\documentclass[preprint,12pt]{elsarticle}

\usepackage{algorithm}
\usepackage{algpseudocode}
\usepackage{amsfonts}
\usepackage{amsmath}
\usepackage{amssymb}
\usepackage[american]{babel}
\usepackage{booktabs}
\usepackage[justification=centering]{caption}
\usepackage[gen]{eurosym}
\usepackage{float}
\usepackage[bottom]{footmisc}
\usepackage[acronym,nohypertypes={acronym,notation}]{glossaries}
\usepackage{graphicx}
\usepackage[hidelinks]{hyperref}
\usepackage{listings}
\usepackage{pdfpages}
\usepackage[figuresright]{rotating}
\usepackage{soul}
\usepackage{xcolor}
\usepackage{xspace}

\newcommand{\toolname}{\texttt{RETINA}\xspace}

\newacronym{CA}{CA}{\textit{Central Authorities}}
\newacronym{DHT}{DHT}{\textit{Distributed Hash Table}}
\newacronym{DL}{DL}{\textit{Distributed Ledger}}
\newacronym{OPEX}{OPEX}{\textit{Operational Expenditure}}
\newacronym{TEE}{TEE}{\textit{Trusted Execution Environment}}
\newacronym{WoT}{WoT}{\textit{Web of Trust}}
\newacronym{PBFT}{PBFT}{\textit{Practical Byzantine Fault Tolerance}}
\newacronym{PKI}{PKI}{\textit{Public Key Infrastructure}}
\newacronym{PoA}{PoA}{\textit{Proof-of-Authority}}
\newacronym{PoW}{PoW}{\textit{Proof-of-Work}}
\newacronym{Pk}{Pk}{\textit{Public key}}
\newacronym{Sk}{Sk}{\textit{Secret key}}

\journal{Sustainable Energy, Grids and Networks}

\begin{document}

\begin{frontmatter}



\title{RETINA: Distributed and Secure Trust Management for Smart Grid Applications and Energy Trading}


\author[1]{Vaios Boulgouras\corref{cor1}}
\author[1]{Thodoris Ioannidis}
\author[2]{Ilias Politis}
\author[1]{Christos Xenakis}
\cortext[cor1]{Corresponding author}
\affiliation[1]{organization={Secure Systems Laboratory, University of Piraeus}, country={Greece}}
\affiliation[2]{organization={Industrial Systems Institute, Research Centre ``ATHENA''}, country={Greece}}

\begin{abstract}
The rapid adoption of smart grids demands robust security and efficiency measures due to their critical role in delivering electricity and their potential for customer-oriented benefits. This paper presents an innovative framework, named \toolname, which provides a resilient and secure energy trading mechanism within smart grid systems. \toolname tackles the inherent security and infrastructure challenges in smart grids by establishing a trust-based security layer and facilitating energy transactions through blockchain technology. Our proposed solution integrates Public Key Infrastructure (PKI) and the Web of Trust (WoT) concepts, promoting decentralized communication channels and robust key management. We further introduce a smart contract-based energy trading mechanism that factors in trust, distance, and energy type (green or non-green) in cost calculation. The utility and robustness of \toolname have been validated in a virtualized testbed environment with 500 nodes, demonstrating superior performance in terms of scalability and resilience compared to the existing WoT scheme. Furthermore, \toolname successfully enables a secure and efficient energy trading scheme, promoting the use of renewable energy sources. Future enhancements will include application to a realistic smart grid deployment and the integration of additional functionalities. This groundbreaking solution has the potential to revolutionize the smart grid ecosystem, addressing its current limitations and propelling the industry towards a future of advanced and secure energy exchange.
\end{abstract}



\begin{keyword}
Smart grids \sep Trust Management \sep Blockchain \sep Security \sep Energy Trading
\end{keyword}

\end{frontmatter}


\section{Introduction}
\label{Introduction}
Smart grids have become a key focus for delivering electricity in a more efficient and eco-friendly manner. Recent technological advancements, such as client-oriented metering mobile apps and provider-oriented data aggregation software, have contributed to this trend~\cite{farmanbar2019widespread}. These advancements offer environmental benefits and financial advantages to consumers, prosumers (those who both produce and consume electricity), and energy providers by reducing operational expenses \gls{OPEX}~\cite{strielkowski2019internet}. As critical national infrastructures~\cite{gunduz2020cyber}, smart grids require continuous operation and integrity protection, making it crucial to implement state-of-the-art security measures.

Robust security measures are essential for smart grids due to their criticality. Establishing a strong security foundation for financial transactions and secure operation of smart applications supporting grid functionalities is mandatory~\cite{tkachenko2019assessment}. As smart grids rely on distributed computing, such as smart meters, information exchange over power lines or wireless channels is susceptible to security threats. To prevent and mitigate these threats, security operators propose solutions utilizing advanced technologies. However, smart grids face technical challenges related to infrastructure security and network architecture that must be addressed to fully realize their potential and enable energy exchange among prosumers.

First, the use of certificate authorities \gls{CA} and nodes in smart grid networks can create single points of failure~\cite{al2021multi}, compromising system safety and functionality. If the CA fails, communication among smart grid nodes may be delayed or suspended, significantly impacting performance and reliability.

Second, while decentralized solutions based on blockchain technology are gaining popularity for their potential to enhance security and transparency in smart grid systems, they can also be resource-intensive and contradict smart grid principles. The maintenance of such solutions can consume significant electricity~\cite{lin2020blockchain}.

Third, there are concerns about the operational cost of energy trading solutions built on the Ethereum network, which currently incurs high transaction fees. This inhibits the adoption of such solutions by providers and customers, reducing their financial benefits.

Lastly, scalability is critical for widespread energy trading adoption among prosumers in smart grids. Existing energy trading solutions primarily focus on small-scale electricity networks, such as microgrids~\cite{fotis2022scalability}. Scaling up poses significant challenges due to increased complexity and resource requirements.

One widely adopted solution for establishing trust between grid components is the Web of Trust (WoT)~\ref{weboftrust}. In the context of smart grids, WoT operates through cryptographic authentication and authorization. Each grid entity is assigned a digital identity, typically in the form of a digital certificate or public key infrastructure (PKI) mechanism, to verify the authenticity and integrity of exchanged data. WoT employs digital signatures to ensure that only trusted entities participate in communication, protecting data during transit.

However, WoT also has shortcomings. The process of assigning and managing digital certificates can be vulnerable to attacks, undermining the trust foundation. Managing a large number of digital identities and certificates becomes complex and resource-intensive as the grid expands. Proper key management, including rotation, revocation, and protection against compromise, is crucial. Efficient mechanisms for trust revocation are necessary to address compromises and unauthorized access, ensuring grid security. Finally, WoT suffers from disturbing the required balance between centralization and decentralization which is important to avoid single points of failure while maintaining effective management and coordination. Addressing these challenges involves robust security protocols, effective key management, standardized practices and collaboration among stakeholders.

In addition to the security challenges posed by the distributed nature of smart grids, energy trading among grid participants has also sprouted along the smart grids. With the smart grid's ability to perform simple computational processes on smart meters, which are essentially IoT devices, the role of energy prosumers has evolved, facilitating the ability to trade energy between two parties over the grid~\cite{sirojan2019embedded}. Energy exchange infrastructures can emerge based on this concept, which is pivotal in enabling modern electric grids to reach their full potential. By aiming for eco-friendly solutions and financial benefits for users, energy trading adds value to consumers compared to legacy grids. Additionally, energy trading complements other solutions, such as pumped storage hydro-power infrastructures~\cite{abdelshafy2020optimized}, that aim to minimize battery usage for storage purposes by instantly distributing and consuming electricity to areas in need. Reduced use of batteries and easy access to green sources of electricity helps decrease the environmental footprint. Moreover, consumers have a variety of available sources offering energy, allowing them to choose the most cost-effective one. By adopting energy trading, the incidence of blackouts due to catastrophic failures in power plants can be significantly reduced.

This paper introduces a framework that establishes the basis for a secure and resilient energy trading mechanism, enabling the secure and efficient operation of smart applications within smart grids. Energy exchange necessitates a robust underlying architecture that adheres to essential standards to facilitate the seamless and secure execution of associated actions. In the absence of those standards, untrusted entities can appear, posing a risk of dishonest trading and potentially leading to market price manipulation and disruption of smart grid operations. In smart grid ecosystems, the concepts of demand and supply play crucial roles. Demand represents the requirement for additional electricity, while supply refers to the surplus electricity that can be provided to the grid. These two concepts are intertwined in a way that they complement each other, offering an opportunity to fulfill the needs of a smart grid without solely relying on the traditional provider. Blockchain technology is employed to facilitate this, leveraging a dedicated smart contract and functionalities that empower participants to place \textit{``buy''} or \textit{``sell''} orders. Blockchain enables transparent and secure energy transactions within the smart grid ecosystem.

More precisely, the paper focuses on developing a distributed security platform called \toolname, which aims to provide prosumer-oriented smart grid environments with robust key management and authentication capabilities. This platform enables independent and decentralized communication channels within the smart grid ecosystem. In addition, the proposed solution incorporates a smart contract-based energy trading mechanism that considers trust, distance, and the type of energy (green or non-green) as attributes in the cost calculation. By integrating trust and energy awareness, \toolname enhances current smart grid implementations. One of the key features of \toolname is its ability to ensure business continuity even with mitigation failures and realized risks, allowing participating nodes to continue their operations securely. Scalability is another vital aspect addressed in the design of \toolname, recognizing the dynamic and evolving nature of smart grids~\cite{venayagamoorthy2011dynamic}. With this in mind, \toolname is designed to be a swift and secure solution that facilitates secure bidirectional communication among smart grid participants. To establish a strong security foundation for \toolname's energy trading mechanism, an innovative hybrid approach combining the \gls{PKI} and \gls{WoT} concepts is employed. This hybrid solution is further augmented by the utilization of Hyperledger Fabric, an advanced blockchain framework, enabling feeless and expeditious transactions. Through the incorporation of Hyperledger Fabric, a decentralized and distributed ledger technology, trust is upheld among network participants while mitigating the risk of malicious actions. Through the energy trading mechanism, prosumers engage in bartering, incentivizing the consumption of energy generated from renewable sources. Overall, the proposed \toolname platform integrates security, scalability, and energy awareness to enhance the functionality and efficiency of smart grids, promoting the adoption of renewable energy sources within the ecosystem.

\medskip \noindent In summary, the paper makes the following main contributions:

\begin{itemize}
\item It introduces \toolname, a novel hybrid approach combining \gls{PKI} and \gls{WoT} architectures on Hyperledger Fabric, enhancing trust through endorsements, attestation, and the security features of blockchain.
\item It overcomes challenges related to resource-friendly trust management and enables fee-less transactions by incorporating a permissioned ledger, providing an efficient solution for energy trading processes.
\item It integrates key exchange, trust management, and energy trading, enabling decentralized participation in green energy production and consumption through smart contracts and a continuous trust mechanism among smart grid entities.
\item The outcomes of \toolname are quantified and validated using a virtualized experimental testbed, demonstrating its superiority over existing trust management solutions for smart grids.
\end{itemize}

The remainder of this paper is organized as follows. In Section~\ref{sec:related-work}, a literature review related to \toolname is presented, focusing on infrastructure security and current energy exchange schemes. The paper's major innovations are also summarized in this section. Section~\ref{sec:architecure} provides a detailed description of the proposed solution's architecture and functions. In Section~\ref{sec:kmevaluation}, a performance evaluation of the proposed framework is conducted, comparing its computation and delay efficiency to state-of-the-art solutions from the literature. Section~\ref{sec:energytrading} focuses on the energy trading aspect of the proposed solution, highlighting the applicability of \toolname. The security risks of \toolname and their impact are discussed in Section~\ref{sec:security}. Section~\ref{sec:critic} critically appraises the proposed framework, discussing the advantages of the hybrid trust establishment scheme. Finally, Section~\ref{sec:conclusion} concludes the paper.

\section{Background}
\label{sec:background}
 In this section the necessary foundational knowledge required to comprehend the key concepts and technologies associated with RETINA are presented.

\subsection{Energy Trading}
Energy trading among prosumers is a dynamic process that operates on the principles of supply and demand, where these entities, who both generate and consume electricity, actively engage in buying and selling electricity based on their specific production and consumption patterns. In this market, prosumers have the opportunity to optimize their energy usage and potentially reap financial benefits by participating as either buyers or sellers, depending on their current energy requirements.

To initiate the energy trading process, prosumers place orders specifying their desired quantity of energy to be bought or sold. These orders reflect their individual needs and preferences, taking into account factors such as their electricity generation capacity, consumption patterns, and any surplus or deficit in their energy supply. Through a transparent and efficient marketplace, these orders are matched with corresponding offers from other prosumers, facilitating the exchange of electricity between parties. By actively participating in energy trading, prosumers can effectively manage their energy resources and optimize their overall energy consumption. For instance, a prosumer with excess energy generation beyond their own consumption needs can choose to sell the surplus to another prosumer who requires additional electricity. Conversely, a prosumer with higher energy consumption demands than their own generation capacity can purchase the needed energy from a willing seller. This mutual exchange of electricity enables prosumers to balance their energy needs, enhance grid stability, and contribute to the overall efficiency and sustainability of the energy system.

Energy trading among prosumers provides a decentralized and market-driven approach to electricity exchange, allowing participants to leverage their individual generation and consumption capabilities. By actively engaging in buying and selling electricity, prosumers play an integral role in shaping the energy landscape, fostering renewable energy adoption, and fostering a more efficient and sustainable energy ecosystem.

\subsection{Public Key Infrastructure}
\gls{PKI} is employed as a fundamental mechanism for creating and signing certificates. PKI serves as a trusted framework that facilitates secure communication and authentication within a smart grid. Each smart meter is equipped with unique cryptographic keys and digital certificates. These certificates serve as a form of identity verification and endorsement for the smart meters. When smart meters interact with each other, they can present their certificates as a means of mutual authentication. By verifying the digital signatures of the certificates using the public keys stored on the blockchain, the smart meters can confirm the identity and trustworthiness of their counterparts.

The utilization of PKI in this context enables a robust and secure endorsement process for smart meters. It ensures that only authorized and trusted devices are involved in energy trading transactions. By relying on the cryptographic capabilities of PKI, the integrity and confidentiality of communication between smart meters are protected, safeguarding against unauthorized access, tampering, or impersonation.

\subsection{Web of Trust}
The utilization of \gls{PKI} enables the establishment of a Web of Trust, which facilitates the operation of the system without the need for a central authority. PKI serves as a foundational framework for creating and managing trust relationships among the smart meters in a decentralized manner. Through \gls{PKI}, a \gls{WoT} is formed among the smart meters, where each smart meter can verify the certificates of other smart meters based on the trustworthiness of the signing authorities. By building and maintaining a network of trusted relationships, smart meters can operate autonomously and securely without relying on a central authority for every transaction or communication.

In this \gls{WoT}, smart meters can validate the digital signatures of certificates using the corresponding public keys stored in their local trust repositories. This process allows them to verify the authenticity of the certificates and the identities of other participating smart meters. The decentralized nature of the \gls{WoT} enables smart meters to establish and manage trust relationships among themselves. As smart meters interact and engage in energy trading processes, they can evaluate the trustworthiness of their counterparts based on the reputation of the signing authorities and the history of successful interactions. By leveraging \gls{PKI} to create a Web of Trust, the energy trading system achieves a resilient and secure operation without relying on a central authority. This decentralized trust model ensures that only trusted smart meters are involved in transactions, mitigating the risks of malicious activity and unauthorized access.

\subsection{Blockchain and Distributed Ledgers}
The blockchain technology, specifically the Hyperledger Fabric platform, is employed to establish two immutable and transparent ledgers that record the endorsed certificates. Each smart meter's signed certificate, representing the endorsement of its identity and trustworthiness, is stored in one of those distributed ledgers. The blockchain ensures the integrity and tamper-resistance of the endorsements, as any modification or tampering attempts would be immediately detectable through consensus mechanisms and cryptographic hashes. The second ledger, utilized for trading orders, facilitates the exchange of energy among prosumers. It serves as a decentralized platform where participants can place their orders to buy or sell electricity based on their production and consumption needs. By leveraging blockchain and distributed ledgers, RETINA provides several key benefits. Firstly, the immutability of the blockchain ensures the integrity of the stored certificates and trade orders, and prevents unauthorized modifications, enhancing the overall security of the system. Secondly, the distributed nature of the ledger eliminates the reliance on a centralized authority, enabling decentralized and autonomous energy trading. Thirdly, the transparency of the ledger allows participants to verify the validity of certificates and track the history of transactions, promoting trust and accountability.

The separation of the two ledgers in RETINA ensures a clear distinction between the storage of certificates (endorsements) and the trading orders. This design choice allows for efficient management and retrieval of endorsements while providing a dedicated platform for energy trading operations. The certificates ledger serves as a foundation for establishing trust among smart meters, while the trading orders ledger enables the secure and transparent exchange of electricity among prosumers.

\subsection{Threat Model}
In the context of RETINA and its energy trading system, it is important to consider a threat model that includes potential malicious actors attempting to manipulate the price of electricity. Malicious actors who may disseminate false or misleading information about energy production or consumption to create a false perception of scarcity or abundance are considered. This can lead to market participants making uninformed decisions and potentially manipulating the electricity price. Furthermore, malicious actors may attempt to compromise the integrity or accuracy of smart meters deployed within the network. By tampering with the measurement or reporting capabilities of these devices, they can manipulate the energy consumption or production data, leading to distorted market conditions and price manipulation.

\section{Architecture}
\label{sec:architecure}

The architecture of the proposed framework follows a ``bottom-up'' approach, beginning with analyzing the smart grid architecture and the role of blockchain. The goal is to establish trust among the network elements and to create a secure and decentralized environment. To achieve this, a hybrid approach that combines the concepts of \gls{PKI} and \gls{WoT} is employed, augmented with blockchain. The \gls{PKI} infrastructure is leveraged to incorporate certificates, which attest to the trustworthiness of individual network nodes. However, instead of relying solely on certificates issued by a centralized certificate authority, the proposed architecture promotes the creation of a Web of Trust among all network nodes. This means that the entire network endorses and recognizes certificates rather than a single authority. This distributed approach ensures the network's resilience if certificate authorities (empowered nodes) become unavailable. In addition, the blockchain technology is crucial in enhancing the security of the smart grid architecture. It provides an immutable and transparent ledger to store the certificates securely. By logging the certificates on the blockchain, their integrity and authenticity can be verified by any participant in a tamper-resistant manner.

The capabilities of \toolname are further extended by incorporating an energy trading mechanism. This mechanism is built on the secure and trusted environment established through the \gls{WoT}, reflected in the ledger entries. Like the trust establishment process, blockchain ensures the immutability and transparency of energy production and consumption activities. These activities are logged on the blockchain, enabling accurate and verifiable customer billing. By integrating the secure and trusted environment of \toolname with the energy trading mechanism supported by blockchain, the proposed framework provides a robust and transparent infrastructure for smart grid operations. It leverages the inherent characteristics of blockchain, such as immutability and transparency, to enhance the security and reliability of the smart grid architecture, ultimately benefiting both consumers and energy providers.

\subsection{RETINA Smart Grid and Blockchain Architecture}

\begin{figure}[t]
\centering
\includegraphics[width=0.8\columnwidth]{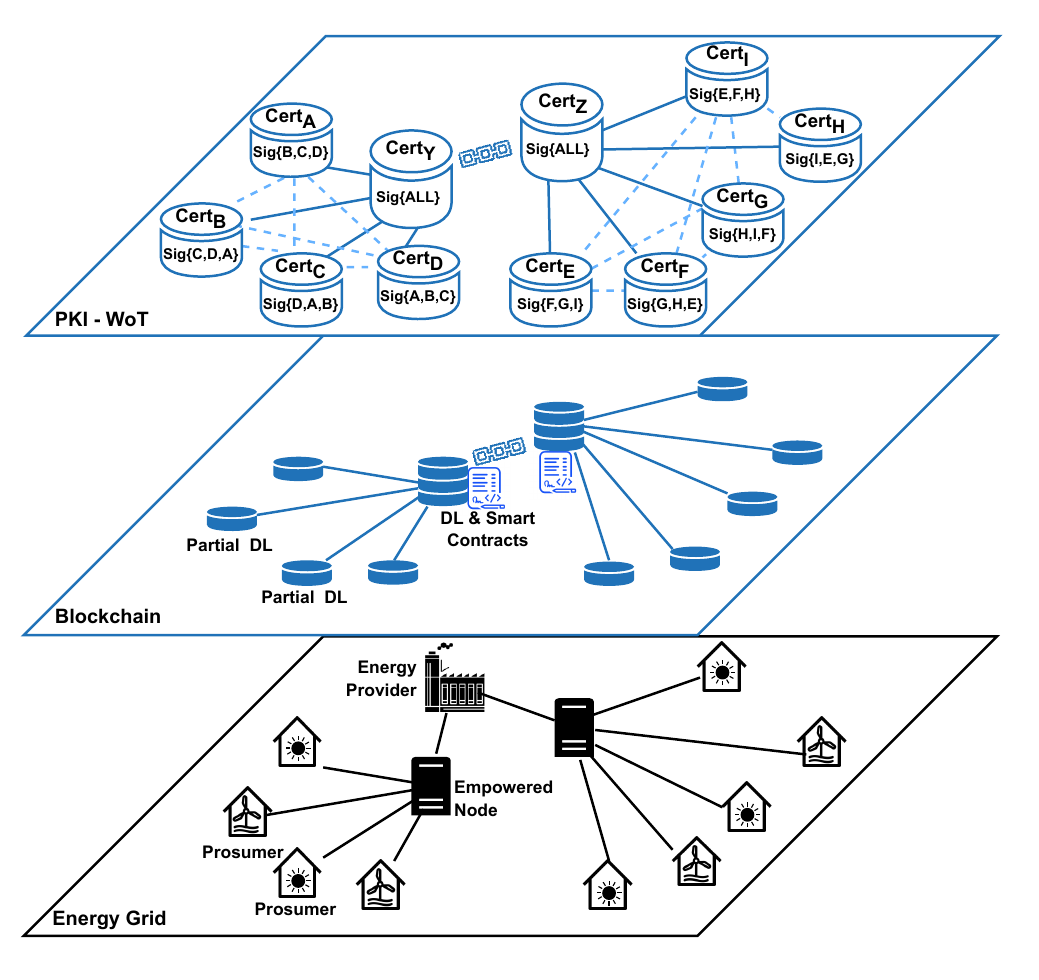 }
\caption{\toolname as an overlay framework, where network participants possess their own certificates, signed by nodes that endorse them}
\label{fig:Arch}
\end{figure}

The proposed smart grid infrastructure, as illustrated in \autoref{fig:Arch}, constitutes of three main components: $(i)$~energy provider, $(ii)$~empowered nodes, and $(iii)$~prosumers' smart meters. The energy provider includes various entities such as energy plants, administration systems, and billing systems. They play a pivotal role in managing and providing energy to the grid. The empowered nodes are strategically positioned in the network to ensure its robustness and mitigate the risk of a single point of failure. Each empowered node defines a neighborhood representing a distinct administrative area within the smart grid. By compartmentalizing the network, administrative responsibilities and resources are distributed among multiple empowered nodes, enhancing the resilience and security of the overall system. Finally, the prosumers' smart meters are key elements in the smart grid infrastructure. They enable measuring and monitoring energy consumption and production at the individual prosumer level. These smart meters interact with the empowered nodes and participate in the secure communication channels established within the framework.

\toolname is depicted as an overlay framework on top of the energy grid in \autoref{fig:Arch}. It utilizes blockchain to provide a secure infrastructure to establish communication channels and gateways. These channels facilitate inter-neighborhood connections, enabling access to a larger market with multiple electricity buying and selling offers. By leveraging the blockchain, \toolname ensures the integrity and transparency of the communication infrastructure. Further, the \gls{PKI} concept is employed on top of the blockchain to facilitate the issuance and management of personal certificates for each smart meter. These certificates serve as a means to establish trust and accumulate endorsements from other participants in the grid. This process creates a \gls{WoT}, where data exchange can occur securely with trusted entities. One of the advantages of this infrastructure is its resilience in the face of empowered node failures. Even if an empowered node becomes unavailable, the grid can continue its operations due to the trust relationships established among the network participants. These relationships are reflected in the \gls{DL}, ensuring the continuity and reliability of the smart grid.

\begin{figure}[t]
\center
\includegraphics[width=0.8\columnwidth]{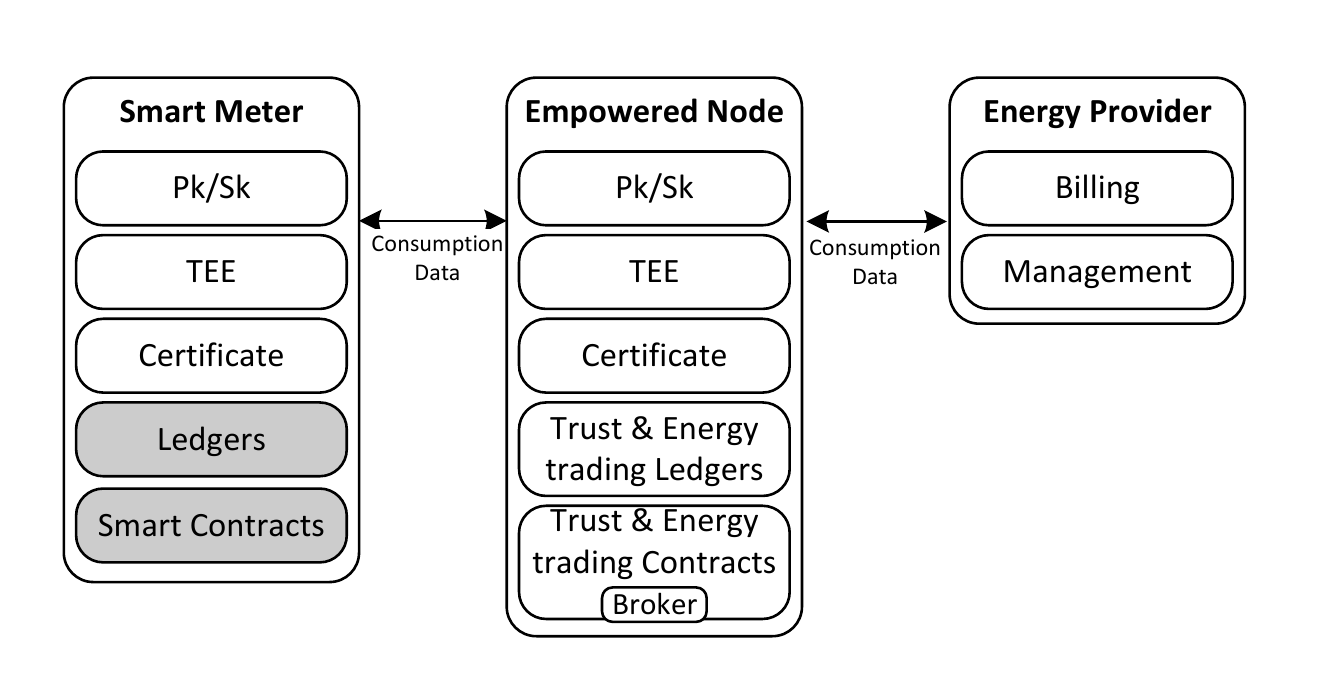}
\caption{Resources and operations within the \toolname network components}
\label{fig:NetComp}
\end{figure}

As shown in \autoref{fig:NetComp}, each smart meter in the smart grid system possesses a Secret/Public (S\textsubscript{k}/P\textsubscript{k}) key pair, along with its associated certificate. To ensure the security of cryptographic operations and protect the secret key, a secure \gls{TEE} is utilized within each smart meter. Additionally, each smart meter maintains a partial copy of the trust management ledger. This ledger contains relevant information about the trust relationships and endorsements among the network participants. Having a partial copy of the ledger enables smart meters to assume the corresponding duties and responsibilities in case an empowered node becomes compromised or unavailable. This feature enhances the resilience and continuity of the system. In essence, the empowered nodes play a paramount role in the blockchain infrastructure. They host the ledgers and smart contracts necessary for the operation of the smart grid system. They are also responsible for resource-intensive operations such as consensus, transaction processing, and executing smart contracts. By distributing these tasks among multiple empowered nodes, the system achieves higher scalability and resilience. Finally, the energy provider, depicted in the diagram, receives information from the empowered nodes for billing and management purposes. It utilizes the data collected from the smart meters and processed by the empowered nodes to calculate and manage energy consumption, billing, and other administrative tasks within the smart grid system.

\begin{figure}[t]
\includegraphics[width=0.7\columnwidth]{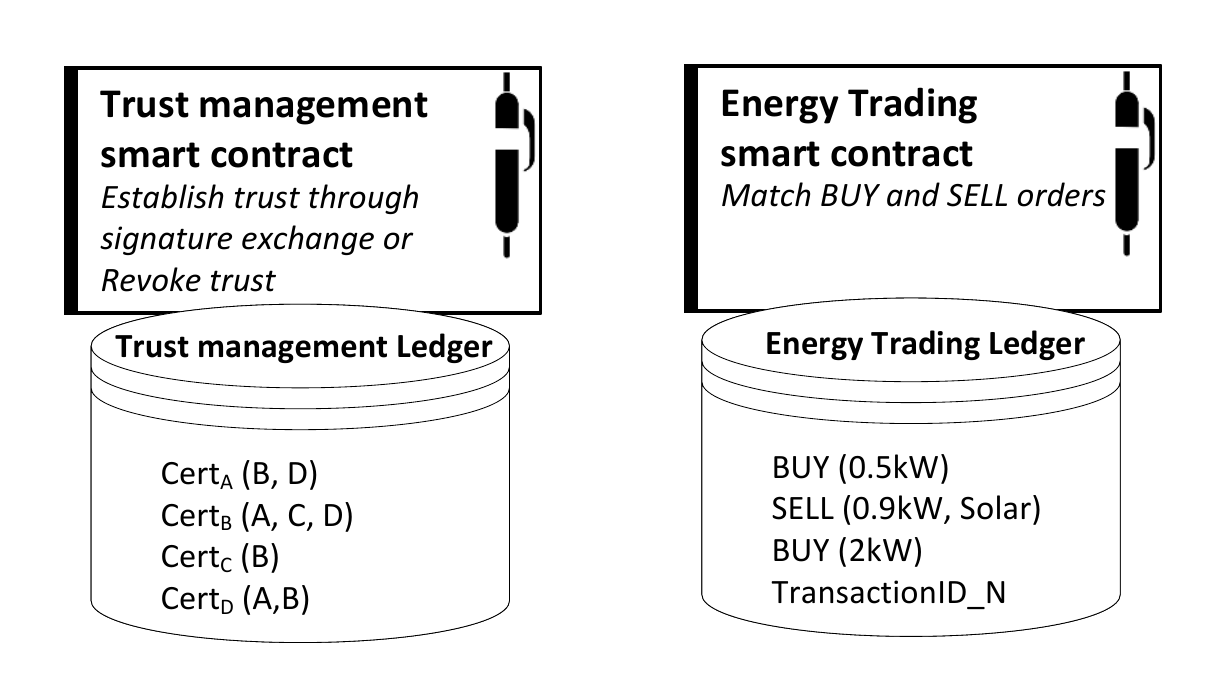}
\centering
\caption{\toolname smart contracts and Ledgers residing on the empowered nodes}
\label{fig:smart-contracts}
\end{figure}

The blockchain possesses vital characteristics that ensure robust security, accountability, and non-repudiation. Moreover, its immutability guarantees the creation of indisputable records. The operational functionalities of the empowered nodes encompass the execution of smart contracts and the maintenance of up-to-date ledgers. As a decentralized node certificate storage platform, the blockchain facilitates energy exchange, where two smart contracts are deployed, and corresponding ledgers are formed, as depicted in \autoref{fig:smart-contracts}. To establish a secure foundation for the smart grid, the ledger is consistently updated whenever a new trust relationship is established, or an existing one is revoked. \autoref{fig:smart-contracts} presents a snapshot of the Trust Ledger, illustrating the bidirectional trust relationship between smart meter $A$ and smart meters $B$ and $D$. The smart contract securely maintains this trust relationship, which automates and safeguards energy transactions for the energy trading mechanism of \toolname, encompassing ``buy'' and ``sell'' orders. The outcomes of completed transactions are accurately recorded on the ledger, ensuring transparency and accountability.

Selecting a consensus algorithm for a blockchain infrastructure is critical, as it significantly impacts system performance and financial implications. In this scheme, the \gls{PoA} consensus algorithm \cite{de2018pbft} has been chosen for utilization by the empowered nodes to achieve consensus and record information on the ledgers. \gls{PoA} is advantageous due to its minimal computational resource requirements, resulting in low energy consumption during operation. The \gls{PoA} consensus mechanism relies on a limited number of validators to approve transactions for ledger entry. This enables frequent updates to the blockchain, reduces the time between block creation, and facilitates the processing of numerous transactions within short timeframes. The scalability of \gls{PoA} is particularly beneficial for smart grids, where the network size may vary and potentially increase substantially. In \toolname, the verification of blocks and transactions is carried out by the empowered nodes, who assume the responsibility of maintaining the blockchain infrastructure.

\subsection{Key Management and Authentication}

The security of the \toolname platform rests on two fundamental pillars: $(i)$~key management operations and $(ii)$~authentication of network entities. These pillars exclude untrusted entities from all network communications and energy trading operations, effectively mitigating the risk of fraudulent activities. The following functionalities are instrumental in realizing this secure scheme.

\subsubsection{New Smart Meter Joining the Smart Grid}

\begin{figure}[t]
\includegraphics[width=0.6\columnwidth]{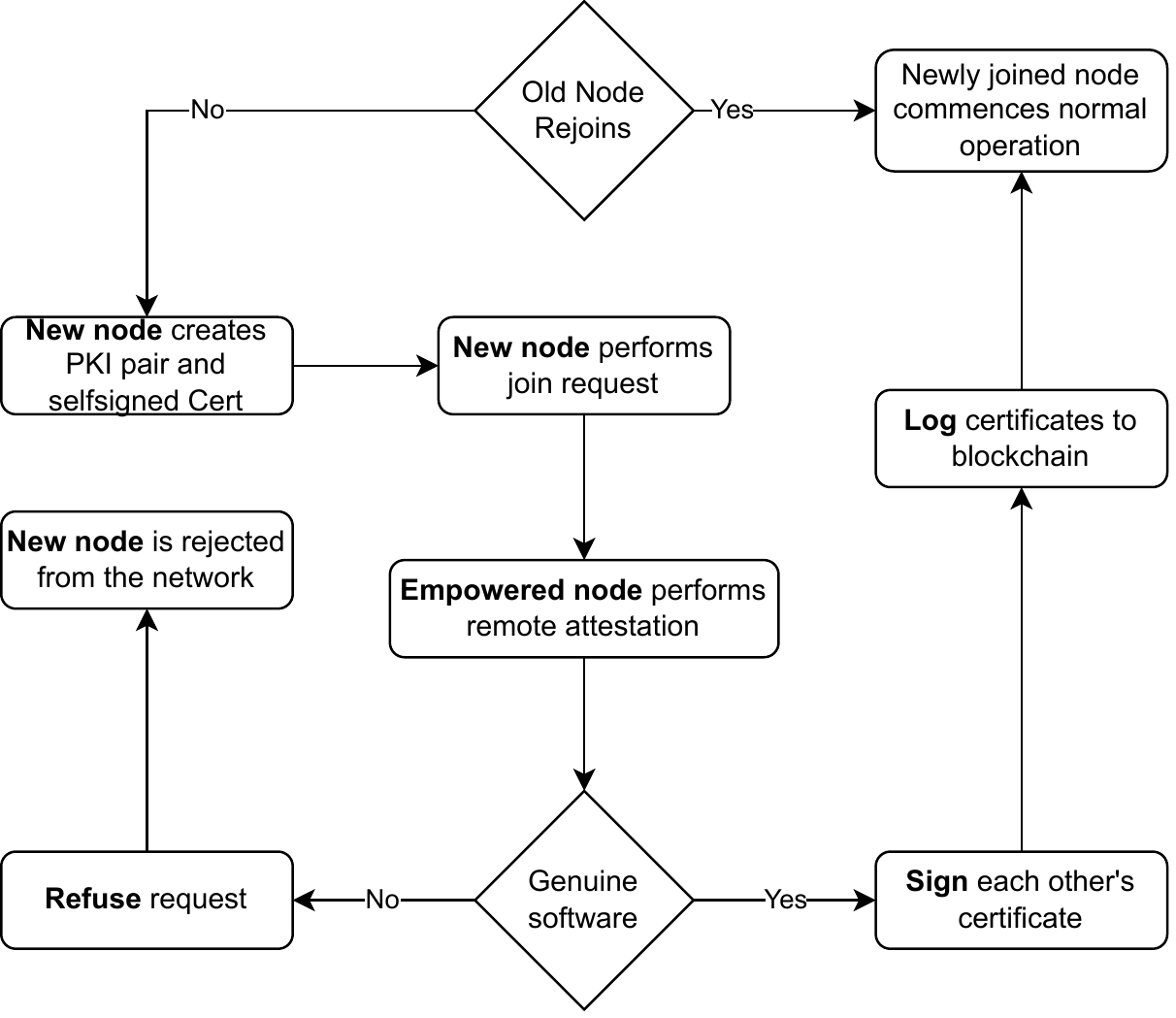}
\centering
\caption{Node join operation}
\label{fig:node-join-operation}
\end{figure}

The process followed by a new node to join the network is illustrated in \autoref{fig:node-join-operation}. In order for a new node to join a smart grid that incorporates the \toolname authentication scheme, it must acquire a signature from at least one empowered node with a valid certificate that is already part of the network, as shown in \autoref{lst:certificate-format}. When a node intends to join the network for the first time, it generates a key pair consisting of a \textit{Public key} ($P_k$), a \textit{Secret key} ($S_k$), and a corresponding self-signed certificate. Subsequently, the new node initiates a request to join the network. A remote attestation procedure is then conducted by the empowered node responsible for the respective neighborhood. This attestation verifies that the candidate node is utilizing provider-approved software and has not been tampered with or subjected to unauthorized alterations. If the remote attestation yields a positive result, the empowered node acting as an introducer signs the certificate of the new node. This signed certificate is logged in the Distributed Ledger (DL) along with the corresponding identification number of the new node, denoted as $ID_n$ (where $n\in[1,N]$, and $N$ represents the network size). In \autoref{lst:certificate-format}, the certificate format is depicted. The \textit{Public key} ($P_k$) is publicly available for all-access, followed by its owner's address (Node~$A$). The first signature accumulated originates from the self-signing step, which precedes the endorsement from an empowered node. Lastly, a signature from Node~$B$ follows, establishing a trust relationship between Nodes~$A$~and~$B$.

\begin{lstlisting}[numbers=none, caption={Format of the certificate}, label={lst:certificate-format}]
Pk 2048R/678455A3 2022-02-07 [expires: 2023-06-07] 
uid Node A <nodea@retina.org> 
sig 962789D1 2023-02-07 Node A <nodea@retina.org> 
sig F4V7287Z 2023-02-07 Empowered <emp@retina.org> 
sig A55CDCC1 2023-02-07 Node B <nodeb@retina.org>
\end{lstlisting}

\begin{figure}[t]
\includegraphics[width=0.5\textwidth]{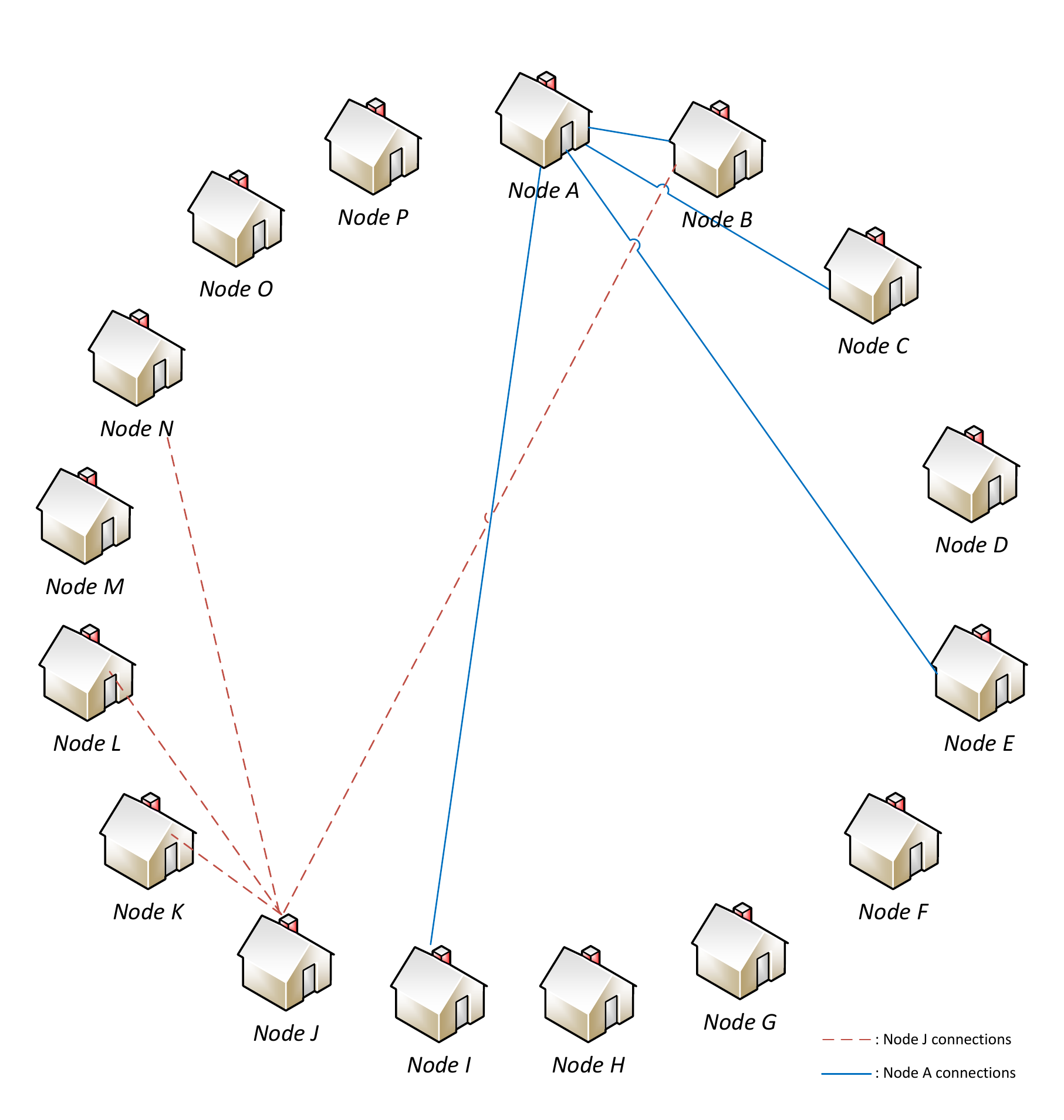}
\centering
\caption{A random snapshot of the process of exchanging signatures, depicting the connections that Nodes $A$ and $J$ make with the nodes that reside on the $2^{N}$ network positions}
\label{fig:exchanges-signatures}
\end{figure}

The new node engages in certificate signature exchanges with the nodes residing in the $2^x$ network positions, where $0 \leq x \leq N/4$, as depicted in \autoref{fig:exchanges-signatures} for Nodes $A$ and $J$. This approach is not employed arbitrarily but is inspired by the demonstrated effectiveness of \gls{DHT} routing algorithms in establishing communication channels among nodes that are not necessarily directly connected~\cite{maymounkov2002kademlia, stoica2001chord}. This eliminates the need for a new node to exchange and store signatures from all other network nodes. In contrast, in a traditional \gls{WoT} network, each node must exchange signatures with all other participants. The utilization of \gls{DHT}-based routing algorithms simplifies the process and reduces the signature exchange requirements for new nodes, allowing them to establish trust relationships with a subset of nodes within the network.

The newly signed certificates are stored locally on the smart meter and the empowered node. To avoid unnecessary storage, simple network nodes maintain only the portion of the ledger corresponding to the endorsements they have given and received. In contrast, empowered nodes possess the complete version of the ledger, including the entire network's trust relationships. Partial ledgers maintained by simple nodes are not actively used during normal operations but are stored and updated. This storage approach ensures that simple nodes do not need to store and manage the entire ledger, reducing their storage requirements. However, they retain the information related to their endorsements and received certificates, enabling them to validate trust relationships when required. If all empowered nodes in a neighborhood are compromised, simple nodes have the potential to assume the elevated responsibilities and role of an empowered node. This ensures the continuity of network operations and the preservation of trust within the neighborhood, even in the face of compromised empowered nodes.

In addition to new nodes, previous network participants who may have gone offline have the potential to rejoin the smart grid, as depicted in \autoref{fig:node-join-operation}. In such cases, if the certificate previously used by the participant is still valid and has not been revoked or expired, no further action is required. The participant can use its existing $ID_n$ and the corresponding certificate already stored on the \gls{DL}. However, if the old certificate is no longer valid, the participant must repeat the aforementioned process of joining the network. To ensure the seamless reintegration of participants who have left the smart grid network and to enable participants with expired or revoked certificates to regain access, a comprehensive procedure is implemented. This procedure involves the generation of a new key pair, obtaining a signature from an authorized node, undergoing remote attestation, and acquiring a newly signed certificate that is recorded on the distributed ledger (DL). By following this process, the smart grid guarantees that participants with valid previous certificates can easily rejoin the network, while also ensuring that those with expired or revoked certificates must go through the necessary steps to regain network access.

\subsubsection{Establishing Trust Relationships}
\label{sec:establishing-trust-relationships}

In a smart grid ecosystem, devices inherently execute applications that require communication among the network's participants. To ensure the smooth execution of these applications, a reliable trust system is necessary. This is achieved by leveraging the initial signature(s) acquired by a new node from the empowered nodes that introduced it to the network. As smart applications are utilized, communication is attempted with other nodes, resulting in an exchange of certificate signing among the network's entities. As mentioned earlier, every new node joining the network, as well as nodes rejoining the smart grid, possesses a certificate signed by one or more empowered nodes. Using this certificate, additional signatures are acquired from common nodes residing in the $2^x$ positions (as indicated by the last signature in \autoref{lst:certificate-format}). A smart contract is employed to facilitate this process. For instance, if Node~$A$ has been introduced to the network by Empowered Node~$S$, the smart contract will utilize Node~$A$'s certificate endorsement by Node~$S$ to introduce itself to other nodes that Node~$S$ has also introduced. This introduction lets the nodes exchange signatures by signing each other's certificates. Normal communication within the smart grid, driven by the operational requirements of smart applications, relies on the trust established through these mechanisms. The trust serves as a foundation for establishing the desired relationships among network participants, allowing for effective and secure communication.

In scenarios where specific empowered nodes (i.e., introducers) are unavailable, and there is no direct trust path between two nodes (i.e., they have not previously signed each other's certificates), the smart contract initiates a process of indirect trust establishment. To provide a clearer understanding of the procedure followed by the smart contract in establishing trust relationships, \autoref{fig:trust-establishment} is provided. The following example illustrates the actions taken when Node~$A$ aims to establish a trust relationship and communicate with Node~$B$:

\begin{figure}[t]
\includegraphics[width=0.6\columnwidth]{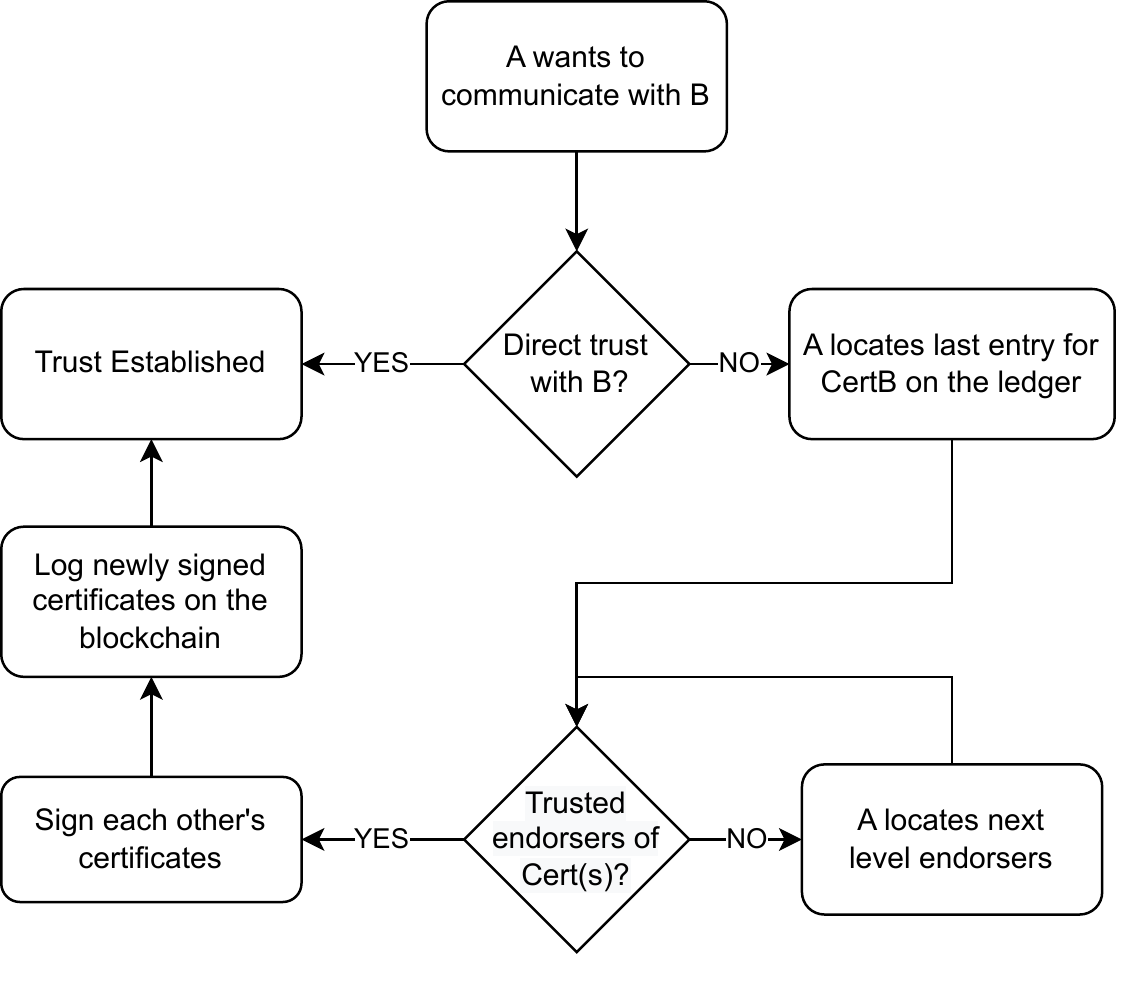}
\centering
\caption{Trust establishment}
\label{fig:trust-establishment}
\end{figure}


\medskip \noindent \textbf{\textit{Step 1.}} Node~$A$ first checks its own certificate to see if it has an existing endorsement from Node~$B$. If such an endorsement exists, the process is completed at this point, and Node~$A$ can establish a trust relationship with Node~$B$.

\medskip \noindent \textbf{\textit{Step 2.}} If Node~$A$ does not find an endorsement from Node~$B$ in its own certificate, a \textit{Certificate Lookup} procedure is initiated. This procedure aims to find the certificate of Node~$B$, denoted as $Cert_{B}$.

\medskip \noindent \textbf{\textit{Step 3.}} Node~$A$ examines the trustworthiness of the endorsers of $Cert_{B}$. If any of the endorsers of $Cert_{B}$ are trusted by Node~$A$, such as Node~$C$, this information is communicated to Node~$B$. Both Node~$A$ and Node~$B$ then sign each other's certificates, log the new trust relationship on the Trust ledger, and conclude the procedure. In this case, only one intermediate entity, Node~$C$, is involved in establishing the trust relationship between Node~$A$ and Node~$B$.

\medskip \noindent \textbf{\textit{Step 4.}} If no trusted endorsers of Node~$B$ are found in Step~3, the Certificate Lookup procedure continues by exploring the signatures collected by the certificates of endorsers of Node~$B$. For example, unlike Step~3, Node~$A$ does not initially trust Node~$C$. However, while searching for a trusted entity in the signatures of the nodes' certificates that have endorsed Node~$B$, Node~$A$ discovers that Node~$D$ has signed the certificate of Node~$C$. This finding establishes a trust path to Node~$B$ through Node~$D$. Therefore, the formed chain of trust, in this case, would be $A \Rightarrow D \Rightarrow C \Rightarrow B$.


\medskip In scenarios where Node $D$ initially lacks trust from Node $A$, Node $A$ can actively search for trusted endorsers of Node $D$ and expand the trust path accordingly. This iterative process continues until a trusted entity is discovered or until a predefined maximum chain length is reached. It is important to note that the performance of the trust establishment procedure, including delays and success probability, can be influenced by the chosen maximum chain length within the \toolname system.

The maximum chain length determines the number of intermediate entities or endorsers that can be explored in the search for a trusted path between nodes. A longer chain length increases the likelihood of finding a trusted path as more intermediate entities are considered. However, it also introduces additional delays in the trust establishment process, as each step necessitates further verification and validation of the endorsers. Conversely, a shorter chain length reduces delays but may limit the probability of finding a trusted path in certain scenarios.

Determining the optimal maximum chain length requires striking a balance between the desired success rate of trust establishment and the acceptable delay in the process. This decision should consider the specific requirements and constraints of the smart grid ecosystem, including network size, the trustworthiness of entities, and performance considerations. Evaluating and optimizing the maximum chain length in \toolname is crucial to achieve efficient trust establishment while meeting the security requirements of the smart grid network.

\subsubsection{Certificate Lookup}

The Trust \gls{DL} employed in the \toolname blockchain implementation is distributed across all network participants. Each participant stores their established trust relationships locally, while the empowered nodes maintain a complete copy of the Trust \gls{DL}. The primary purpose of the Trust \gls{DL} is to facilitate the certificate lookup procedure, which occurs on the empowered nodes when a direct trust relationship is not present but is desired. To perform the certificate lookup procedure, the empowered nodes utilize the process outlined in Section~\ref{sec:establishing-trust-relationships}. This process identifies the nodes that should be considered when finding a trust path between two unrelated nodes. As previously explained, trust paths can be established indirectly through multiple intermediate nodes. This, combined with the fact that the Trust \gls{DL} is processed locally on the empowered nodes, offers several advantages:

\begin{itemize}
\item \toolname's swiftness is not limited by network constraints that may be presented due to bandwidth limitations or increased overhead because of heavy traffic.
\item The network load will not be burdened with extra traffic due to the certificate lookup operation, which will be executed locally.
\end{itemize}

\begin{figure}[t]
\includegraphics[width=0.7\textwidth]{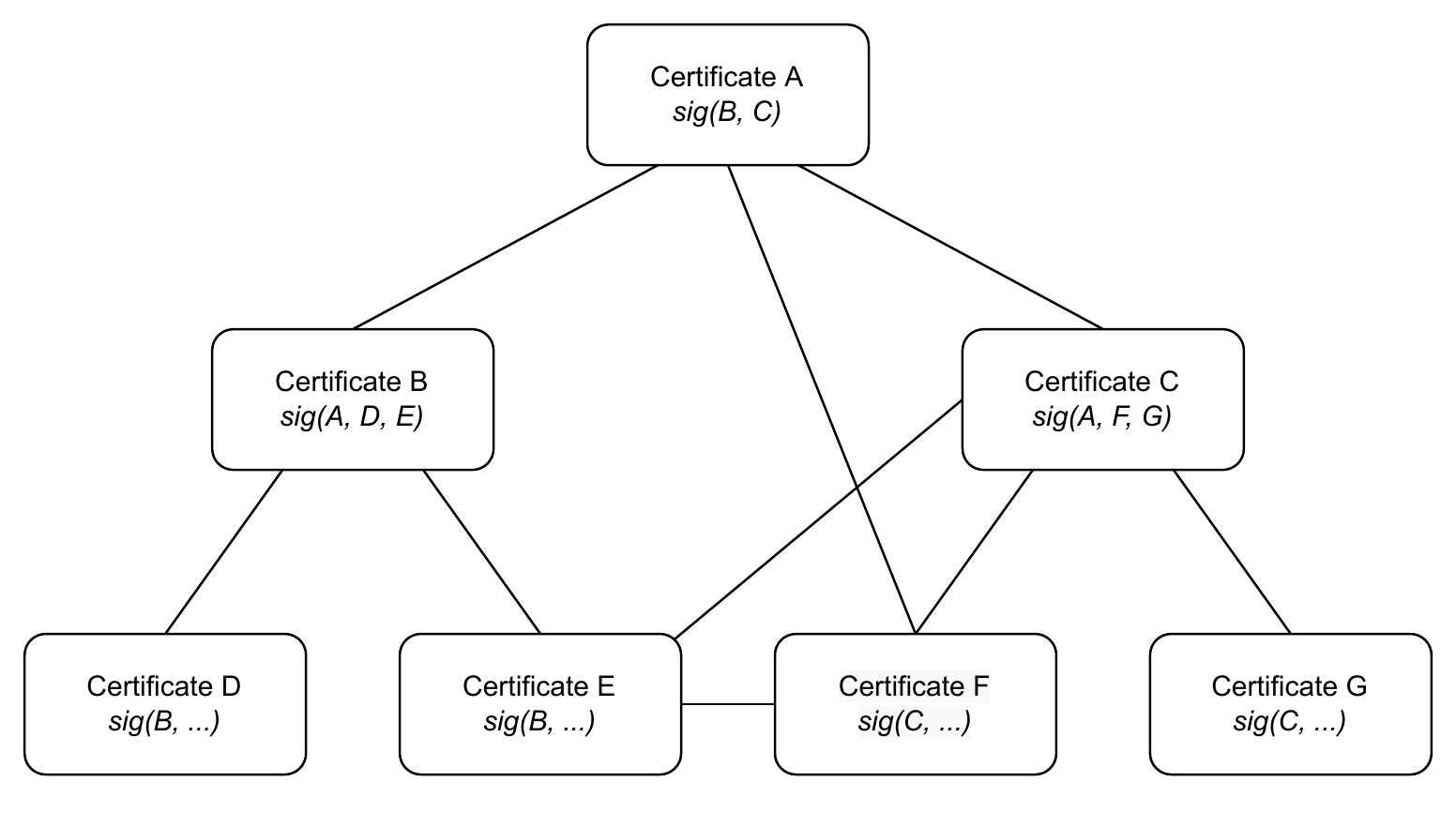}
\centering
\caption{Snapshot of the Trust Paths that have been formed among seven network participants}
\label{fig:trust-paths}
\end{figure}

\autoref{fig:trust-paths} represents a visual depiction of an instance found in every node's locally stored Trust \gls{DL}. This simplified representation illustrates an already-formed \gls{WoT} among seven nodes. Each node's certificate contains multiple accumulated signatures, establishing new trust relationships. In this instance, Node~$A$ has exchanged signatures with Nodes $B$, $C$, and $F$, as indicated by the signatures in their respective certificates. Likewise, Nodes~$B$ and $C$ have established trust paths with other nodes. Node~$B$ has trust paths with Nodes~$A$, $D$, $E$, and Node~$C$ has trust paths with nodes $A$, $E$, $F$, and $G$. This visualization of the trust relationships showcases the interconnected nature of the trust paths. It highlights how trust can be established indirectly through the involvement of intermediate nodes. The Trust \gls{DL} stored on each node enables the identification and utilization of these trust paths to facilitate secure interactions and transactions within the blockchain network. Overall, this graphical representation demonstrates the practical application and effectiveness of the Trust \gls{DL} in enabling trust establishment and verification between nodes in the \toolname blockchain implementation.

To better understand the \textit{Lookup Procedure}, an example based on \autoref{fig:trust-paths} is presented. In this scenario Node~$A$ needs to contact Node~$E$. The following actions will be performed:

\medskip \noindent \textbf{\textit{Step 1.}} The initial search focuses on the signatures that $Cert_{A}$ has accumulated in order to determine whether the target Node~$E$ has already signed it or not.

\medskip \noindent \textbf{\textit{Step 2.}} In this scenario Nodes~$A$ and $E$ have not exchanged signatures yet, so a search for trust with one intermediate node commences by examining the endorsers of Node's~$E$ certificate.

\medskip \noindent \textbf{\textit{Step 3.}} The inquiry of Step~2 returns a positive result, as Node~$B$, which is trusted by Node~$A$, is an endorser of $Cert_{E}$. At this point the lookup procedure is concluded successfully and the process to establish a trust relationship by exchanging signatures, as described in the previous section, commences.

\medskip Following the same train of thought, if the target node was at the next level in terms of depth, the procedure's execution would be extended, and an additional intermediate would be included in the formed trust path. Due to the nature of the network's ring-like structure, similar to the one found in \gls{DHT} networks, an iterative-recursive hybrid approach~\cite{kunzmann2005recursive} was utilized in terms of search routing, benefiting the overall performance of the system as showcased in Section~\ref{sec:kmevaluation} below.

\subsubsection{Certificate Revocation}

Node certificates may be revoked for two main reasons. Firstly, a node may voluntarily leave the network and choose to revoke its certificate. Secondly, a node might be considered untrustworthy or potentially malicious. To ensure the integrity of the network, all participating entities undergo periodic remote attestation conducted by empowered nodes. As described by Panos et al.~\cite{panos2014specification}, this attestation process involves generating a hash value from each reviewed node's \gls{TEE}. This hash value is derived from predetermined software components and parameters that should remain unchanged. If most nodes conducting the remote attestation procedure conclude that a particular node's software has been tampered with and could potentially engage in malicious actions, its corresponding certificate is revoked. For instance, in \autoref{fig:trust-paths}, if $Cert_{B}$ is revoked, the associated signatures in $Cert_{A}$, $Cert_{D}$, and $Cert_{E}$ would be removed. This revocation process also applies to empowered nodes to safeguard the smart grid against malicious actors. Irrespective of whether the revoked certificate belongs to an empowered or a common node, the network's operations remain unaffected due to the existing \gls{WoT}, which interconnects all smart grid nodes directly or indirectly. In the aforementioned example, the revocation of $Cert_{B}$ as an intermediate certificate between Nodes~$A$ and $E$ can be compensated by Nodes~$C$ and $F$. Furthermore, the revoked certificate will not be utilized in any future attempts by other nodes to establish trust relationships among themselves.

Once a certificate revocation occurs, the trust relationships the revoked node establishes become unavailable if the node later rejoins the network. In such cases, the procedure for a new smart meter joining the grid must be repeated only after addressing the compromised software that initially led to the node's exclusion from smart grid operations. This entails issuing a new pair of keys and the associated certificate. Conversely, nodes that abruptly depart from the smart grid due to Internet connection issues or technical problems do not need to repeat the join process since their certificate remains valid.

The certificate revocation process is inherently sensitive due to potential abuse, which could lead to denial of service. To mitigate this risk, the remote attestation performed by empowered nodes to network participants is securely protected and inaccessible to users or third parties. This safeguarding is achieved by utilizing \gls{TEE} technology, which masks and shields these operations from malicious actors. The calculation of attestation hashes from the attestee, and the verification process conducted by the attesters occur within this protected environment, rendering them immune to tampering by any entity. Additionally, cryptographic processes related to certificate signing and energy trading rely on TEE utilization to mitigate various risks associated with fraudulent activities.

\section{Performance Evaluation}
\label{sec:kmevaluation}
The proposed framework has undergone rigorous testing and validation across various scenarios, comparing it to state-of-the-art solutions in terms of communication and computational costs during trust establishment. A comparison against Web of Trust (WoT) implementations confirms that RETINA significantly improves the time required to establish trust in scalable smart grid environments and demonstrates superior responsiveness when empowered nodes are lacking. To demonstrate the practicality of RETINA in real-world scenarios, performance evaluation has been extended to validate an energy trading scheme that utilizes RETINA to establish trusted nodes as energy traders. This scheme ensures security guarantees in cases of trust revocation or unauthorized access by malicious traders. The evaluation of the RETINA-based energy trading scheme focuses on examining energy trading price fluctuations resulting from tampering by malicious nodes. Through these evaluations, the framework's reliability, efficiency, and security in trust establishment and energy trading scenarios are demonstrated, providing compelling evidence of its applicability and effectiveness in real-world deployment.

\subsection{RETINA Configuration}

The simulation of the \toolname framework was deployed on a physical machine running \textit{Ubuntu 18.04 LTS}, equipped with an \textit{Intel i7 6700HQ} processor and 32 gigabytes of RAM. The implementation was carried out in Java, utilizing smart contracts for Key management and Energy trading functionalities as the main components. To facilitate the blockchain deployment of \toolname, the Hyperledger Fabric framework was employed with the assistance of the \textit{Hyperledger fabric-gateway-java} dependency. Multiple nodes were simulated to assess the network's performance and responsiveness in various real-life scenarios. Each node was assigned corresponding certificates and wallets for these simulations and evaluations. The certificates adhere to the PGP standard and employ the \texttt{SHA-1} hash function, which generates block hashes. 

\subsection{Communication Cost and Computation Time}

As depicted in Table~\ref{tab:comcom-cost}The evaluation of \toolname primarily centers around its comparison to related works, specifically regarding two key factors: $(i)$~computation time required for signature exchange between two nodes, and $(ii)$~communication costs associated with the data transferred during the mutual endorsement process between network nodes. These evaluation criteria aim to assess the performance and efficiency of \toolname to similar approaches in the field.

\begin{table}[t]
\centering
\caption{Comparison in terms of communication cost and computation time between \toolname and other similar solutions}
\label{tab:comcom-cost}
\resizebox{\columnwidth}{!}{
\begin{tabular}{lcc}
\toprule
\textbf{Frameworks} & \textbf{Computation Time (ms)} & \textbf{Communication Cost (bits)} \\  
\midrule
\texttt{FeneChain}~\cite{li2020blockchain} & 2000 & 540.67 \\
\texttt{SURVIVOR}~\cite{jindal2019survivor} & 117.35 & 448 \\
\texttt{BEST}~\cite{chaudhary2019best} & 35 & 961 \\
\texttt{Secure ET in DRM}~\cite{kumari2020blockchain} & 33 & 968 \\
\texttt{EnergyChain}~\cite{aggarwal2018energychain} & 33 & 576 \\
\toolname & 19.2 & 192 \\
\bottomrule
\end{tabular}
}
\end{table}

To calculate the computation time, we measured the duration of various blockchain-related operations necessary for establishing trust. These operations encompassed the following:

\begin{itemize}
\item \textbf{\textit{VerifyCertificate:}} Use the certificate's \gls{Pk} to decrypt and verify its validity. For \toolname, the average time to decrypt and validate the certificate was $2.7ms$.
\item \textbf{\textit{SignCertificate:}} Use own \gls{Sk} to sign a certificate. In \toolname, the time required to sign the certificate was calculated at $2.7ms$ on average.
\item \textbf{\textit{Addition:}} Making a new entry on the \gls{DL} (\texttt{addToDL}). On average, the time required for this process in \toolname is measured at $1 ms$.
\item \textbf{\textit{One-way hash function (SHA-1):}} Hashing the information pertaining to a new block (\texttt{hashData}). The SHA-1 operation required in \toolname an average of $2.7ms$. 
\item \textbf{\textit{Append:}} Collecting data into a single string, giving it a proper form to introduce it to the blockchain (\texttt{addNonce}). The append operation in \toolname has been timed at $0.5ms$ on average. 
\end{itemize}

We obtain the computation time denoted as $CT$ by summing the average times for each operation. Since the endorsement process for establishing a trust relationship is executed mutually from both ends, the entire process is performed twice, once for each node involved. Therefore, we calculate the total Computation Time as $CT_{total}$ by doubling the computed value of $CT$. This accounts for the bidirectional nature of the endorsement process, ensuring a comprehensive evaluation of the computational efficiency of \toolname in establishing trust between network nodes.
%
\begin{equation*}
\small
CT=\texttt{VerifyCertificate} + \texttt{SignCertificate} + \texttt{addNonce} + \texttt{hashData} + \texttt{addToDL}
\end{equation*}
\begin{equation*}
\small
CT_{total} = 2CT
\end{equation*}

The communication cost ($CC$) in the case of \toolname is evaluated based on the number of bits transferred during the establishment of a trust relationship between two nodes. In this context, we consider the specific components involved in the process:
\begin{itemize}
\item \textbf{\textit{Block Header Size:}} The number of bits processed at the block header, denoted as $Hb$, is calculated as: \textit{Hb = [32 bits Identity + 256 bits previous block hash + 32 bits transactions] = 320 bits}.
\item \textbf{\textit{Hash Output:}} Utilizing the SHA-1 hash function, the output size of the hash, denoted as $Hout$, is determined as \textit{Hout = SHA-1(320 bits + 32 bits nonce + 64 bits padding) = 416 bits}. 
\item \textbf{\textit{SHA-1 Input and Digest Size:}} The total input size for the SHA-1 algorithm is 416 bits, and after digesting, the final output size is 160 bits.
\end{itemize}

Based on these calculations, the total cost of the \gls{PoA} consensus algorithm is determined as \textit{PoA = 32 bits Identity + 160 bits SHA-1 digest = 192 bits}. It is important to note that this communication cost is borne solely by the Empowered nodes due to the nature of the \gls{PoA} consensus algorithm implemented in \toolname.
%
%
\begin{equation*}
\small
CC = \texttt{ID} + \texttt{SHA-1 digest}
\end{equation*}

\subsection{RETINA Compared to Web of Trust} 

An infrastructure based on the \gls{WoT} was implemented, drawing from the descriptions outlined in the OpenPGP standard~\cite{callas2007rfc} and the corresponding official guide~\cite{zimmermann1995official}), as a decentralized trust model. This implementation enables a comparative performance analysis between \toolname and \gls{WoT}, highlighting the distinct advantages of each solution. Specifically, we examine the decentralized performance of \toolname's algorithm in contrast to a traditional \gls{WoT} network, focusing on the time required for $(i)$~introducing a new node to the network and $(ii)$~establishing communication between two network participants. Notably, previous literature has conducted network experiments in similar scenarios involving network sizes ranging from 100 to 500 nodes~\cite{zhang2019novel, yang2019blockchain, selvaraj2022capture}. To ensure comparability with these studies, our performance evaluation encompasses network sizes of 100, 200, 300, 400, and 500 nodes.

\begin{figure}[t]
\includegraphics[width=0.7\columnwidth]{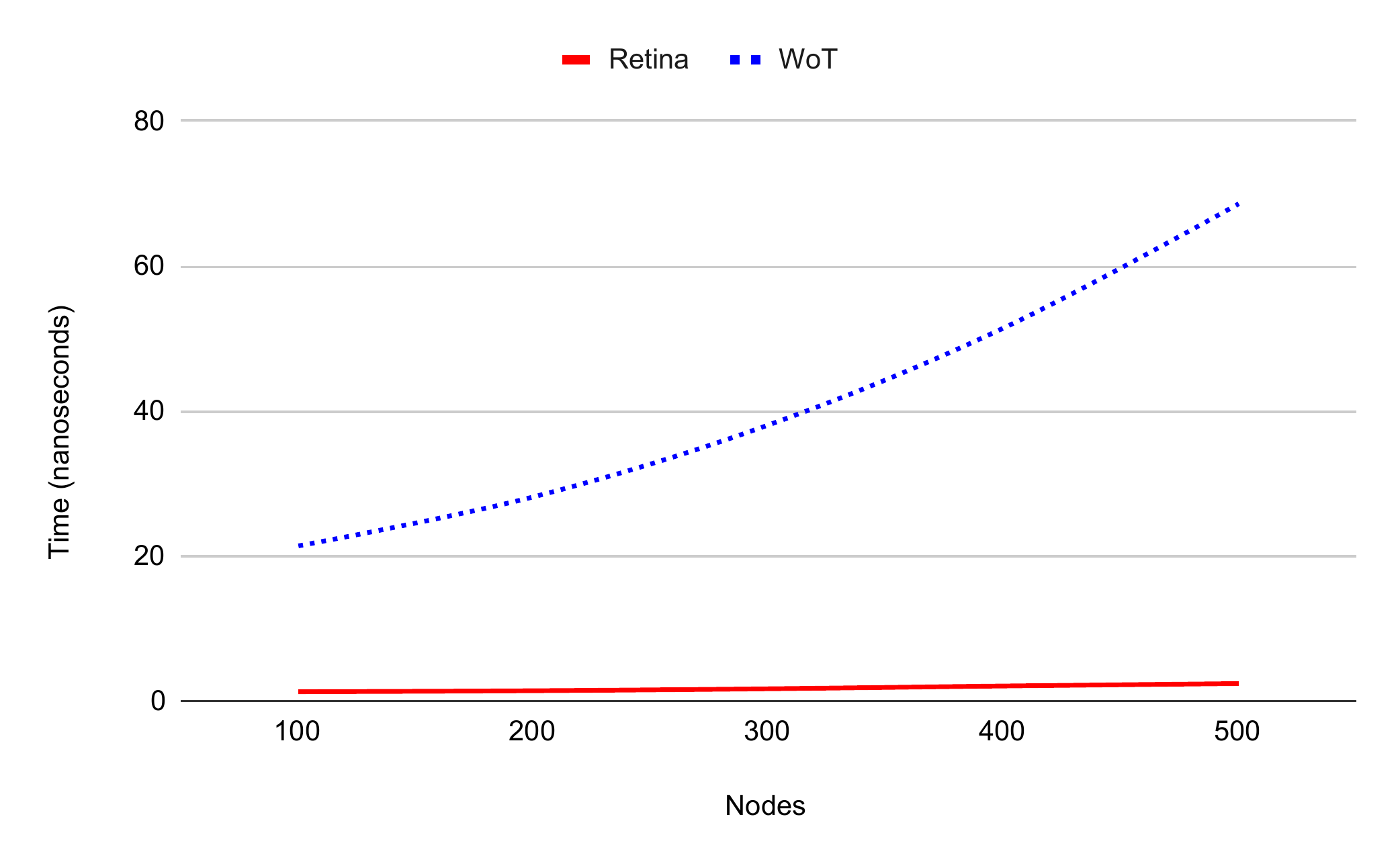}
\centering
\caption{Comparison in terms of time required for a  node to establish a new trust relationship with a random node under \toolname and \gls{WoT} schemes, for different network sizes}
\label{fig:new-node-initialization}
\end{figure}

\autoref{fig:new-node-initialization} demonstrates the substantial performance advantage of \toolname over a traditional \gls{WoT} network when it comes to the initialization of new network nodes, as indicated by the introduction times ranging from $1.2ms$ to $2.7ms$ for \toolname, compared to the considerably longer durations of $21.4ms$ to $144.7ms$ required by a node joining a \gls{WoT} network. This discrepancy in performance can be attributed to the distinct mechanisms employed by each system. In \toolname, a new node only needs to exchange signatures with the nodes occupying the $2^x$ positions, whereas, in a \gls{WoT} network, the new node must introduce itself to all other nodes through a flooding procedure.

\begin{figure}[t]
\includegraphics[width=0.7\columnwidth]{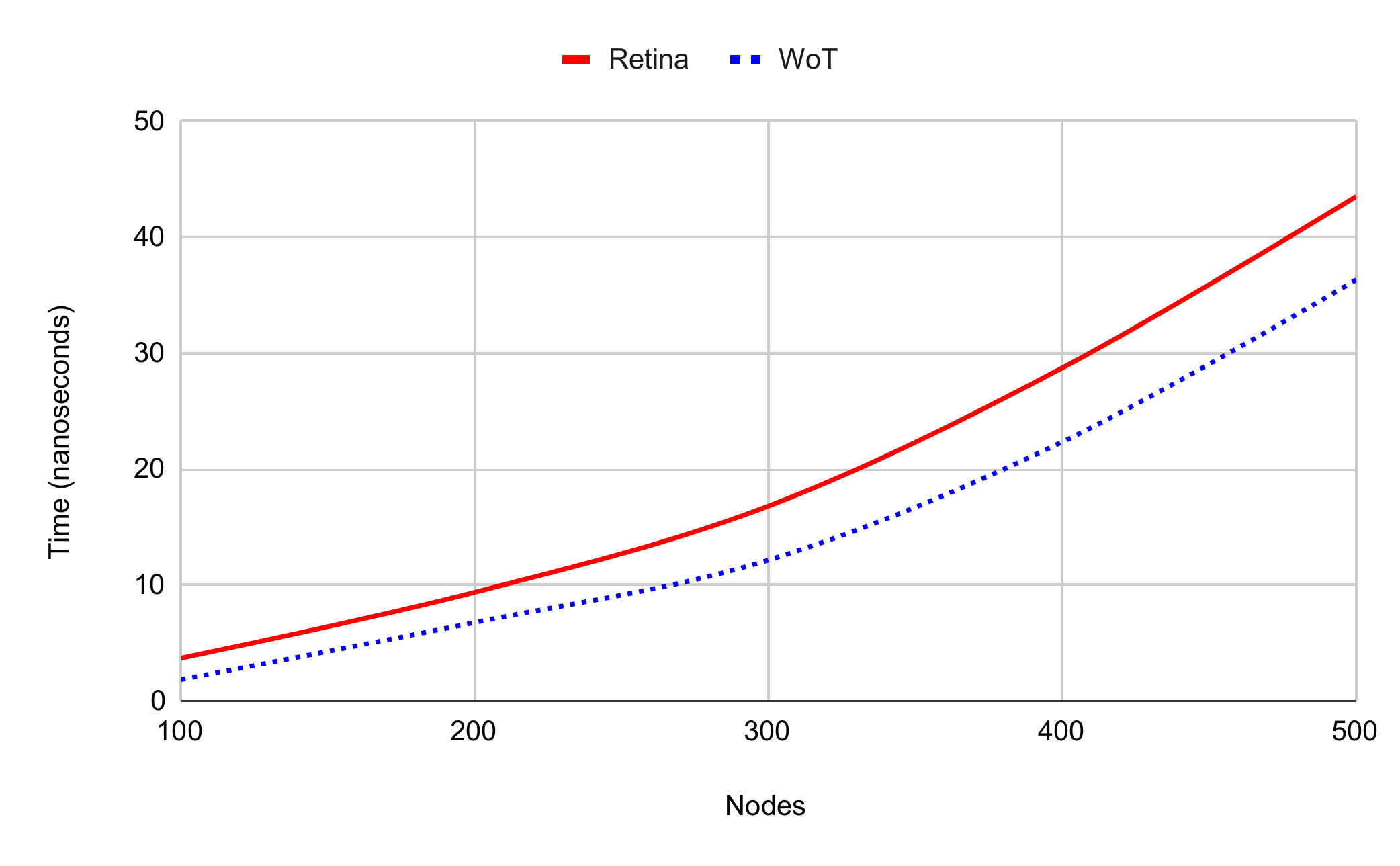}
\centering
\caption{Comparison in terms of time required for a new node to join the network and establish trust among the first tier nodes under \toolname and WoT schemes, for different network sizes}
\label{fig:trust-establishment-delay}
\end{figure}

While \toolname demonstrates superior performance during the introduction phase of new nodes, there is a slight trade-off regarding trust establishment, as illustrated in \autoref{fig:trust-establishment-delay}. In a \gls{WoT} network, a node needs to query its database to locate the endorsement for another node, whereas, in \toolname, this may not always be necessary. If the endorsement is not locally stored on a node, a search is initiated using trust relationships established by other trusted nodes. However, thanks to \toolname's recursive/iterative hybrid search method, the time delay for establishing a trust relationship remains relatively low, comparable to that of a \gls{WoT} node. It is important to note that this metric applies explicitly to a decentralized \toolname scenario, where the empowered node is unavailable. In other cases, \toolname's performance closely aligns with that of a \gls{CA}-centered network, resembling the \gls{WoT} approach, as both involve searching a single database.

\section{Energy Trading}
\label{sec:energytrading}
To comprehensively illustrate the benefits of the proposed framework and demonstrate the practicality of \toolname, we integrated an energy trading mechanism for smart grids. This mechanism empowers all participating entities, predominantly prosumers, to engage in buying or selling the energy they generate within the smart grid ecosystem. By establishing an exchange platform among these entities, we facilitate increased consumption of green energy. This is achieved through the availability of diverse energy sources and the seamless accessibility provided by the trading mechanism. Consequently, integrating \toolname enables a more sustainable and environmentally friendly utilization of energy resources within the smart grid infrastructure.

\subsection{Energy Trading Mechanism}

Within \toolname, every network node possesses a wallet that contains the quantity of produced energy and a corresponding balance. The energy trading mechanism implemented in \toolname leverages the trust relationships formed during the earlier stages of the nodes' life cycle. All entities involved in the exchange processes are considered trusted, indicating that their certificates have received one or more signatures from valid endorsers. This reliance on established trust ensures the integrity and reliability of energy transactions within the network, fostering a secure and efficient energy trading environment.

The ``buy'' and ``sell'' orders within the energy trading mechanism are posted on the blockchain and signed by the entities issuing them. This approach ensures that the orders are openly accessible to all participants in the network, allowing for easy verification of their authenticity. These orders possess specific characteristics, including:

\begin{itemize}
\item \textbf{\textit{kW:}} The amount of energy to trade.
\item \textbf{\textit{Production method ($CO_{2}$ footprint):}} It states whether the power in \textit{kW} referred to in an order has been produced using renewable energy sources or not.
\item \textbf{\textit{Location:}} It refers to the smart grid location, in terms of neighborhood, of the order originator. 
\end{itemize}

Defining the $Price/kW$ in the energy trading mechanism is not deterministic and considers the aforementioned characteristics, which can vary between nodes. This architecture is flexible and can accommodate various pricing models. For the described energy trading scheme, a theoretical pricing model has been chosen, incorporating the listed criteria through the utilization of two weights: $a \in [0.2, 0.5]$ to account for the trading of green energy (where a lower $a$ value corresponds to a higher proportion of green energy being offered), and $b \in [0.2, 0.5]$ to account for the distance between the trading nodes (with a higher $b$ value indicating a greater distance). It should be noted that the specific values of these weights are merely indicative and can be adjusted according to the policies of the energy provider. Thus, the theoretical price assigned to the energy trading contract between two nodes in the grid can be calculated using the following formula, where \texttt{StartingPrice} represents the base price per $kW$.

\begin{equation}
\label{eq:price}
\small
Price=StartingPrice+a(StartingPrice)+b(StartingPrice)
\end{equation}

During periods of high load on the grid's infrastructure, prices within the energy trading mechanism may favor participants who exchange energy with nearby counterparts within the same neighborhood. This approach helps minimize the utilization of network resources, and it can be achieved by setting a minimum value for the \textit{``location proximity''} weight, thereby reducing the corresponding energy cost accordingly. To illustrate this concept, consider the following example, highlighting how the cost of \textit{1kW} of energy can vary based on its specific traits. It is important to note that additional factors, such as the scarcity of electric power during peak hours, can also be considered in the price calculation, adhering to the fundamental principles of supply and demand.

Consider the example of \textit{1kW} of energy (with a base price of \euro{}1/\textit{kW}) produced from renewable energy sources (where $a=0.2$) originating from a node located in a different neighborhood (with $b=0.5$). In this scenario, the indicative price of the energy transaction can be calculated using \autoref{eq:price}, resulting in $Price = 1 + 1 \times 0.2 + 1 \times 0.5 = 1.7$. Therefore, the calculated price for this particular energy transaction would be \euro{}1.7.

Consider now the case of \textit{1kW} of energy (with a base price of \euro{}1/\textit{kW}) generated from renewable energy sources (with $a=0.2$) originating from the same neighborhood (with $b=0.2$). Using the same calculation as in \autoref{eq:price}, the price for this transaction, accounting for the impact of distance, would be $Price = 1 + 1 \times 0.2 + 1 \times 0.2 = 1.4$. Therefore, the calculated price for this specific energy transaction would be \euro{}1.4.

In the energy trading mechanism, the nodes automatically place orders based on the status of their individual energy reserves. A broker, represented by an algorithm running on the smart meters, considers multiple variables to determine whether it is necessary to place either ``sell'' or ``buy'' orders. These variables include the current rates of electricity production and consumption and the availability of battery reserves. By analyzing this data, the broker creates a projection regarding the potential need or surplus of electricity. Moreover, the broker considers behavioral patterns observed in the past, such as increased electricity demand at noon. Based on these considerations, the broker places the corresponding orders on the ledger. 

The energy trading smart contract then takes control, matching ``buy'' and ``sell'' orders that complement each other. The smart contract ensures that the conditions of each order are met before executing the transaction. Additionally, it should be noted that the broker can also receive manual input from the user. This enables the fulfilling of specific needs that may arise and cannot be anticipated solely by the broker's algorithm. Once a transaction is completed, the result is recorded on the ledger, including the updated values of the remaining energy and balance of the node involved in the transaction. This logging process ensures transparency and accountability within the energy trading system.

\begin{algorithm}[t]
\caption{Trading smart contract}
\label{alg:trading-smart-contract}
\scriptsize
\begin{algorithmic}[1]
    \If{$EnergyReserves = HIGH$}
        \State $Action \gets SELL;$
    \ElsIf{$EnergyReserves = LOW$}
        \State $Action \gets BUY;$
    \EndIf
    
    \If{$Action = SELL$}
        \For{$every Buy order$}
            \State $Get StartingPrice;$
            \State $Price = StartingPrice + StartingPrice \times locationProximityWeight;$
            \State \textit{Save OrderID if price is the highest so far;}
        \EndFor
        \State \textit{Sell to highest offer;} \Comment{Provided that minimum expectations set beforehand are met}
        
        \If{\textit{no Buy orders}}
            \State \textit{Create Sell order}
        \EndIf
    \ElsIf{$Action = BUY$}
        \For{$every Sell order$}
            \State $Get StartingPrice;$\Comment{greenEnergyWeight has been factored in}
            \State $Price = StartingPrice + StartingPrice \times locationProximityWeight;$
            \State \textit{Save OrderID if price is the lowest so far;}\Comment{may skip orders concerning non-green energy}
        \EndFor
        \State \textit{Buy lowest offer;} \Comment{Provided that minimum expectations set beforehand are met}
        
        \If{\textit{no Sell orders}}
            \State \textit{Create Buy order}
        \EndIf
    \EndIf
\end{algorithmic}
\end{algorithm}

The trading process is comprehensively outlined in \autoref{alg:trading-smart-contract}, encompassing a series of steps that ensure efficient electricity exchange. Initially, each node determines whether it needs to buy or sell electricity. If a node possesses an excess amount of electricity, it diligently searches for the most favorable ``buy'' offer available at that particular moment. Once the search concludes, the ``sell'' order is meticulously examined to ascertain if it meets any predetermined minimum requirements established by the seller. Subsequently, upon finding a ``buy'' order that satisfies these prerequisites, the transaction is promptly executed. However, if no ``buy'' order meets the seller's specified requirements, a corresponding ``sell'' order is generated with the expectation of being matched by an equally qualified ``buy'' order. The same procedure is reversed when a node intends to purchase electricity. In this scenario, the node strives to identify a matching ``sell'' order and, if unsuccessful, creates a ``buy'' order accordingly.

\subsection{Performance Evaluation of the Trading Mechanism}

The experimental setup for the subsequent evaluation remains consistent with the description provided in Section~\ref{sec:kmevaluation}. Leveraging the pre-established trust relationships, a series of experiments were conducted to showcase the intended functionality of the \toolname energy trading system, enabling nodes to achieve independence from traditional providers. These experiments highlight that price fluctuations align with participants' inclination to buy or sell electricity, per the fundamental principle of supply and demand. To facilitate this demonstration, a simulation environment was created, encompassing a predetermined number of neighborhoods, participating nodes, and their corresponding attributes: $(i)$~individual electricity reserves, $(ii)$~electricity reserve thresholds which when crossed the participant seeks to either buy or sell electricity, and $(iii)$~electricity consumption and production rate.

It is worth noting that to attain more realistic outcomes, randomness was incorporated into the individual electricity reserves, consumption rates, and production rates. This stochastic component was introduced to better simulate real-world scenarios' inherent variability and uncertainties. By including this randomness, the experimental setup aimed to capture the dynamic nature of electricity trading and accurately reflect the diverse characteristics and behaviors of participating nodes.

Three distinct scenarios have been explored, each based on the initial intentions of the participating nodes. These scenarios encompass:

\begin{center}
\textbf{\textit{1$^{st}$ Scenario:}} 75\% Buyers, 25\% Sellers
\end{center}

\begin{center}
\textbf{\textit{2$^{nd}$ Scenario:}} 50\% Buyers, 50\% Sellers
\end{center}

\begin{center}
\textbf{\textit{3$^{rd}$ Scenario:}} 25\% Buyers, 75\% Sellers
\end{center}
    
\begin{table}[t]
\centering
\caption{Energy trading network configurations}
\label{tab:marketplace-network-configurations}
\resizebox{\columnwidth}{!}{
\begin{tabular}{lcc}
\toprule
& \textbf{Network Configuration 1} & \textbf{Network Configuration 2} \\
\midrule
Neighborhoods & 10 & 3 \\
Nodes per Neighborhood & 30 & 100 \\
Market Cycles & 30 & 30 \\
\bottomrule
\end{tabular}
}
\end{table}

For each scenario, two distinct network configurations were employed during the simulation, denoted as \textit{Network Configuration 1} and \textit{Network Configuration 2}. The \textit{``Market Cycle''} concept was established to represent a timeframe within which each network node engaged in electricity production, consumption, and corresponding buying or selling activities based on their energy reserves. \autoref{tab:marketplace-network-configurations} provides an overview of the network configurations utilized for the simulation. In \textit{Network Configuration 1}, the area was divided into ten neighborhoods, each with ten nodes. Conversely, \textit{Network Configuration 2} maintained the same number of total network participants but redistributed the nodes across a smaller number of neighborhoods. By comparing the electricity price reactions in both configurations, the aim was to demonstrate the feasibility and unimpeded interconnectivity between neighborhoods. Furthermore, it was assumed that the provider offered incentives to encourage the consumption of green energy, leading to the definition of specific weight values as follows: $(i)$~\texttt{greenEnergyWeight = 0.2} and $(ii)$~\texttt{locationProximity = 0.5}.

\begin{figure}[t]
\includegraphics[width=0.7\columnwidth]{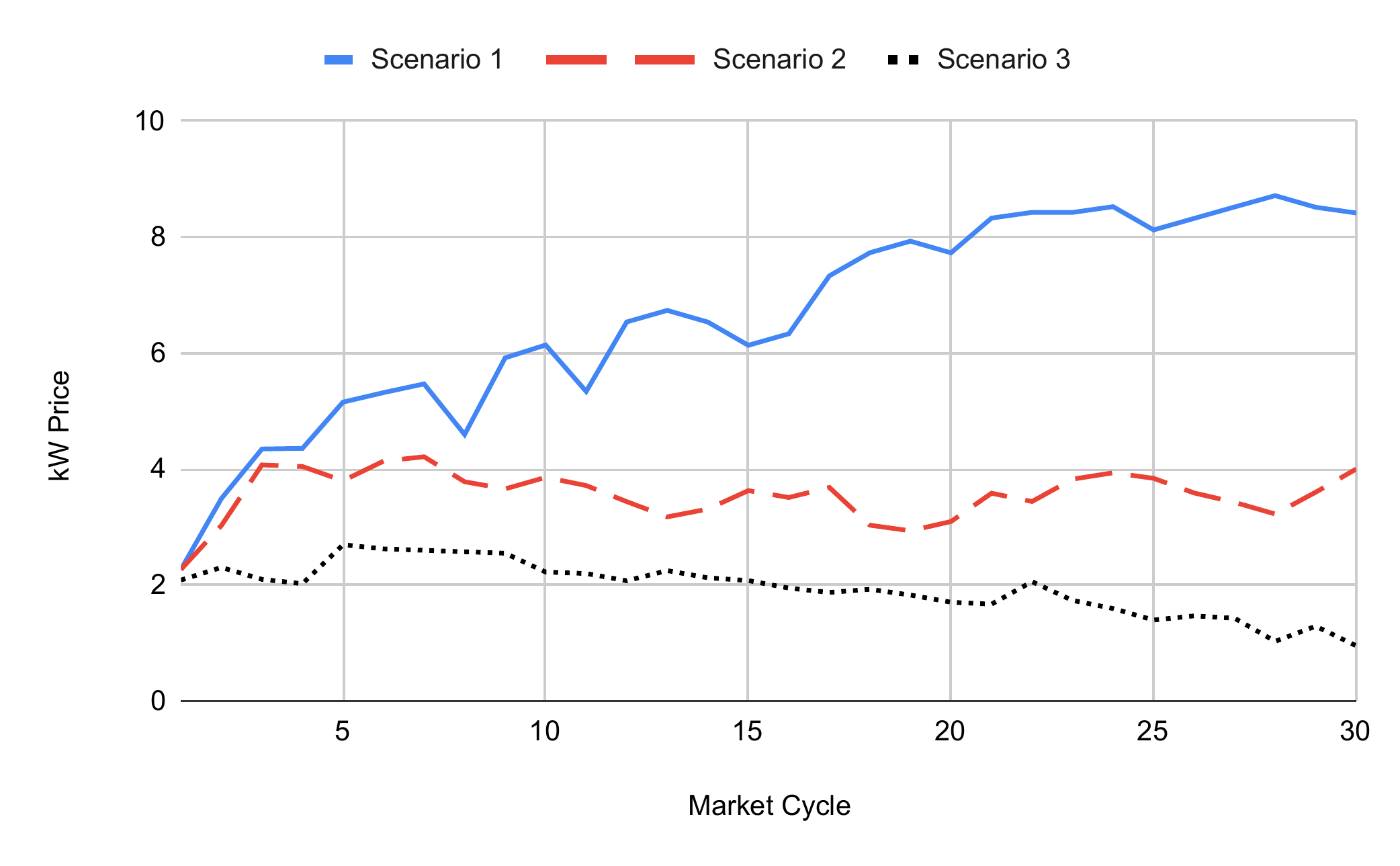}
\centering
\caption{kW price chart -- Network configuration 1}
\label{fig:kW-price-chart-network-configuration-1}
\end{figure}

\begin{figure}[t]
\includegraphics[width=0.7\columnwidth]{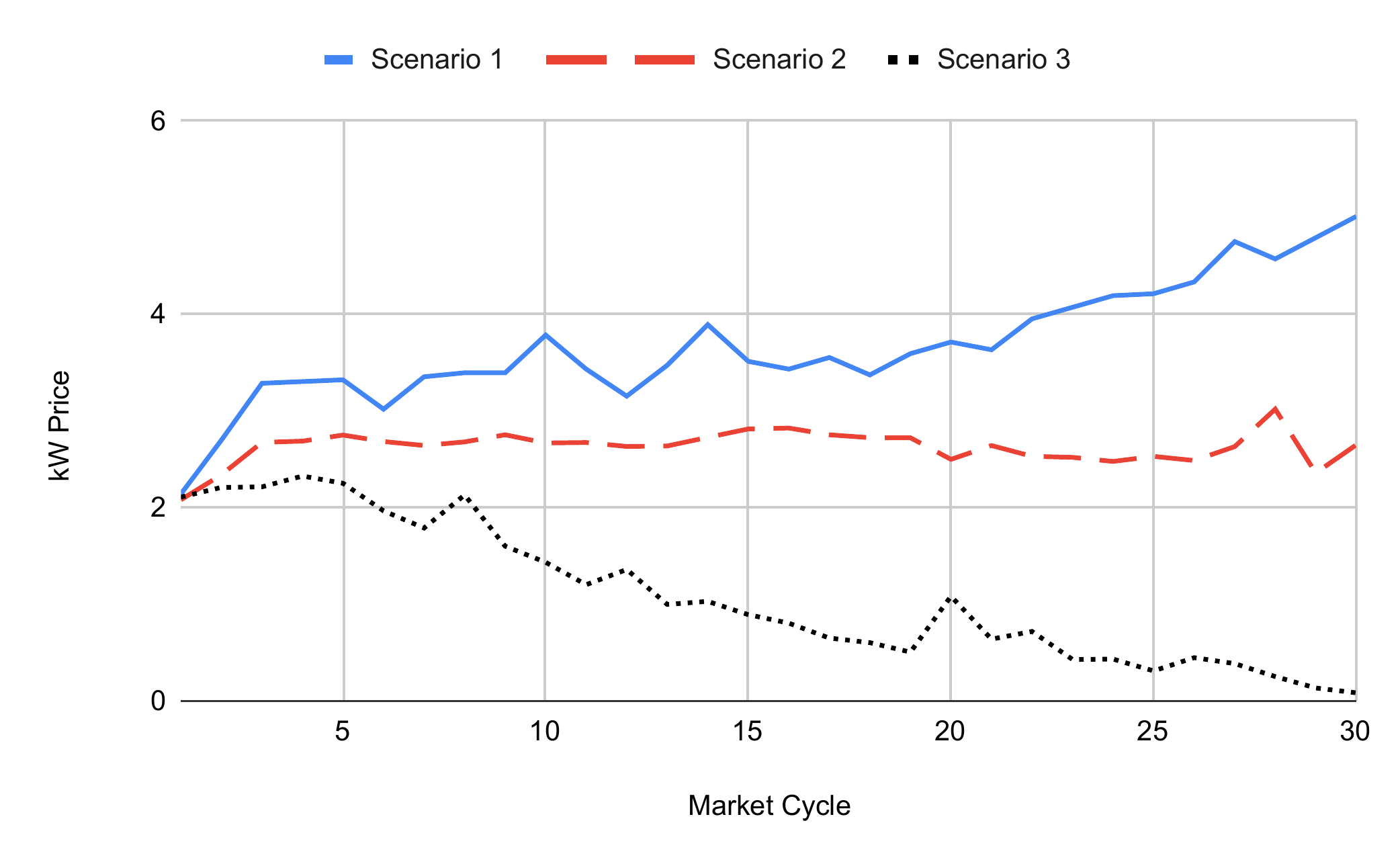}
\centering
\caption{kW price chart -- Network configuration 2}
\label{fig:kW-price-chart-network-configuration-2}
\end{figure}

In both \textit{Network Configuration 1} (\autoref{fig:kW-price-chart-network-configuration-1}) and \textit{Network Configuration 2} (\autoref{fig:kW-price-chart-network-configuration-2}), the expected price fluctuations are observed, aligning with the principles of supply and demand. When the number of buyers exceeds that of sellers, the price increases, while the opposite scenario leads to a decrease in price. The introduction of randomness in the simulation results in a realistic price graph that reflects the electricity needs of the network. In \textit{Network Configuration 1}, where a higher presence of buyers is combined with the majority of buyers and sellers residing in different neighborhoods, resulting in a weight of 0.5 factored into the final price calculation, the price is driven higher compared to \textit{Network Configuration 2}. The brokers successfully execute their roles by considering the nodes' requirements and finding the most suitable Sell or Buy orders. Notably, the results demonstrate that the entire grid is self-sustained, with none of the nodes relying on the Utility for electricity.

\section{Security Analysis}
\label{sec:security}

The subsequent security analysis primarily identifies potential attack scenarios that may arise due to the innovative characteristics inherent in the described architecture. The aim is to assess the system's vulnerability to security threats and devise appropriate mitigation measures. By comprehensively examining the security landscape, it becomes possible to enhance the robustness and resilience of the architecture, ensuring the protection of critical resources and maintaining the integrity of the overall system.

\subsection{DoS Attacks}

The functionality of \toolname heavily relies on the empowered nodes, particularly regarding blockchain operations. While the network operates in a distributed manner without a single point of failure, a coordinated attack targeting these specific nodes could potentially disrupt the entire network. However, robust mitigation measures have been implemented, including establishing trust relationships among entities independent of empowered nodes. In the event of such an attack, the most trusted node(s) within the affected neighborhood would assume an elevated role, temporarily replacing the compromised empowered node. This ensures the continuity of network operations and growth, as demonstrated in the performance evaluation, even in a worst-case scenario. The system's resilience safeguards against potential disruptions and underscores the effectiveness of the implemented mechanisms in addressing security concerns.

\subsection{Energy Trading Privacy}

The privacy of information stored on the ledgers is an inherent characteristic of the permissioned blockchain employed by \toolname. This ensures that entities not participating in the network cannot access the associated data. However, it is important to acknowledge the possibility of a node being compromised and maliciously exploited to leak information outside the smart grid ecosystem. If such a scenario occurs, indicating a failure of the remote attestation procedure, measures are in place to anonymize the stored information using the \gls{Pk}/\gls{Sk} pairs possessed by each node through the utilization of \gls{TEE}. This ensures that external parties cannot obtain readable and usable information unless they are part of the network. Specifically, while the provider can match a node ID to a specific smart meter or empowered node, readable and usable information remains inaccessible to external entities, preserving the data's privacy.

\subsection{Attack of the Clones}

Like the well-known $51\%$ attack and the Byzantine problem, the proof of authority consensus mechanism can be vulnerable~\cite{ekparinya2019attack} if there is an insufficient number of nodes available to validate transactions relative to the network's size. However, this vulnerability can be effectively mitigated by ensuring an adequate number of empowered nodes within the network. Additionally, including seemingly ordinary nodes that are prepared to assume authoritative responsibilities in the event of a security incident or compromise of an empowered node further strengthens the system's resilience. By maintaining a sufficient number of empowered nodes and establishing contingency plans with backup nodes, the potential risks associated with the proof of authority consensus mechanism can be effectively addressed, bolstering the overall security and stability of the network.

\subsection{Energy Trading Manipulation}

The electricity offered within the energy trading mechanism can be regarded as a tradable commodity, accompanied by various associated risks. One such risk involves the potential for exploiting price differentials by purchasing inexpensive electricity during the night and selling it at a higher price during the day. However, several factors contribute to the unlikelihood of such an attack materializing:

\begin{itemize}
\item \textbf{\textit{Continuous Production of New kW:}} Prosumers continuously generating new electricity mitigates the probability of experiencing a shortage on the grid. This ongoing production ensures a relatively stable supply, reducing the feasibility of manipulating prices through artificial scarcity.
\item \textbf{\textit{Price Adaptation:}} In response to an increase in demand, whether during the day or night, the pricing mechanism within the energy trading system adapts accordingly. This dynamic adjustment helps balance supply and demand, limiting opportunities for price manipulation.
\item \textbf{\textit{Optimized Broker Behavior:}} The broker component of the system is designed to learn from the behavior of the energy trading mechanism, enabling it to make informed decisions. By monitoring patterns and identifying nodes that frequently require external assistance for electricity, the broker can anticipate future needs and ensure fulfillment during periods of lower electricity prices.
\end{itemize}

By considering these factors, the energy trading mechanism mitigates the risks associated with opportunistic trading strategies and promotes a fair and efficient marketplace for electricity exchange.

\section{Critical Appraisal}
\label{sec:critic}

In summary, \toolname presents a solution that addresses the challenges outlined in Section~\ref{Introduction}, offering smart grids the means to overcome these obstacles and unlock their full potential by incorporating an energy trading mechanism. This achievement is made possible by strategically integrating several established and proven technologies and concepts, including blockchain, \gls{PKI}, and the \gls{WoT}. By combining these elements, \toolname establishes a robust foundation that synergistically complements one another, providing a comprehensive solution for the smart grid domain.

By implementing \toolname, a notable advancement towards achieving autonomy and independence from centralized solutions within smart grid ecosystems is realized. While a central authority, represented by the provider, establishes policies and configures the network initially, once normal operations commence, smart grids exhibit self-sovereign characteristics and can autonomously perform fundamental functions. \toolname's blockchain infrastructure addresses the limitations often encountered in related literature that employs similar \gls{DL} concepts. This innovative solution minimizes operational costs and optimizes the utilization of computational resources available within smart meters, leading to enhanced efficiency and improved resource management. The integration of \toolname thus marks a significant milestone in progressing towards fully autonomous smart grid ecosystems.

The scalability is a critical aspect supporting the claim of \toolname's ability to enable fully decentralized smart grid ecosystems. The experiments demonstrate that \toolname-based smart grids can maintain their functionality in a decentralized manner, even with networks comprising at least 500 nodes. Given the nature of smart grids, which involve the execution of multiple applications such as the energy trading mechanism, constant and frequent communication among nodes is inevitable. Consequently, communication packets are transmitted within short intervals, resulting in an incremental ``incubation'' time for the network. This inherent characteristic ensures that \toolname can effectively handle the increasing communication demands as the network expands, facilitating the seamless operation of various applications within the smart grid ecosystem. The scalability of \toolname further strengthens its viability and potential for large-scale deployment in real-world smart grid scenarios.

Direct access to collectively produced green energy from smart grid participants holds significant importance for prosumers, liberating them from the limitations of their infrastructures and the provider's capacity to supply electricity. With the implementation of the energy trading mechanism, participants gain access to a vast green energy market, enabling them to buy and sell electricity directly. This arrangement offers substantial financial benefits for all involved parties while contributing to environmental sustainability. By facilitating direct transactions within the smart grid ecosystem, \toolname empowers prosumers to actively participate in the green energy market, fostering a more efficient and environmentally conscious energy consumption landscape.

\section{Related Work}
\label{sec:related-work}

The existing body of research on smart grid security is vast, with numerous works exploring various aspects such as user and device authentication, data privacy and confidentiality, and secure communication channels. To provide a comprehensive understanding of the innovations proposed in this paper, the related work is discussed in two distinct categories: $(i)$~security solutions for smart grids and $(ii)$~security solutions specifically designed for energy trading mechanisms. By examining the related work in these categories, this paper aims to highlight the novel contributions of the proposed framework in addressing the security challenges of both smart grids and energy trading mechanisms. The discussion of previous research provides valuable insights into the existing state of the art, enabling a clear understanding of the advancements and innovations offered by the proposed solution.

\subsection{Smart Grid Security}

The following works focus on the authentication and privacy aspects provided by blockchain technology in the context of IoT and smart grids, which are vital for successfully implementing endeavors such as an energy trading mechanism. Yakubov et al.~\cite{yakubov2018blockchain} propose an authentication solution that combines an x.509 certificates ecosystem with blockchain technology, covering processes like issuance, validation, and revocation of certificates. However, a drawback of this approach is that if the certificate authority loses access to the blockchain and is excluded from the network, a new certificate authority smart contract must be created, completely restructuring the network's certificates. In contrast, \toolname handles the exclusion of central nodes differently, allowing operations to continue based on endorsements among network nodes.

Wang et al.~\cite{wang2019blockchain} present an anonymous authentication and key management scheme for smart grids utilizing edge computing and blockchain technology. Experimental results demonstrate its efficiency and speed. However, a single point of failure is observed in their solution, where if an edge server goes offline, the corresponding nodes' connection to the blockchain network is disrupted. This issue is not present in \toolname, as nodes are independently connected to the blockchain.

Zhang et al.~\cite{zhang2019blockchain} design a decentralized access-control manager that utilizes the \gls{PBFT} consensus algorithm for participant authentication. The network's security relies on the assumption that malicious nodes will not reach a threshold defined by the \gls{PBFT}'s security requirements, which limits scalability. In contrast, \toolname employs remote attestation based on \gls{TEE} to exclude malicious nodes from the network rather than tolerating them.

Finally, Bolgouras et al.~\cite{bolgouras2019distributed} combine the \gls{PKI} and \gls{WoT} concepts into a hybrid solution tailored for microgrids. While innovative and efficient in search speed, this solution falls short compared to \toolname regarding nodes joining and leaving operations. In their approach, when a node goes offline or rejoins the network, a reconstruction of successor nodes where certificates are stored is required, whereas, in \toolname, certificates remain on the \gls{DL} without any alterations.

By leveraging blockchain technology in \toolname, network security is enhanced through the blockchain's transparency, immutability, and accountability properties. \toolname improves detection of malicious actions by auditing the \gls{DL} entries and renders a \gls{DHT} obsolete for information logging purposes.

\subsection{Energy trading}

Park et al.~\cite{park2019market} and Bosco et al.~\cite{bosco2018blockchain} propose energy trading schemes using the Ethereum blockchain infrastructure. However, the high transaction fees associated with the Ethereum network reduce the financial benefits for potential adopters. In contrast, \toolname is built on the Hyperledger platform, utilizing a \gls{PoA} consensus algorithm, which enables transactions without fees.

Aitzhan et al.~\cite{aitzhan2016security} present a decentralized energy trading system using blockchain technology. The authors emphasize the requirement for multiple signatures to execute a transaction. A transaction must be signed by a predetermined number of participating entities in their scheme before being submitted to the blockchain. In \toolname, entities form trust bonds and do not need to sign numerous transactions constantly.

Regarding payments in energy trading, Aitzhan et al.~\cite{aitzhan2016security} utilize the Bitcoin system, which relies on the power-hungry \gls{PoW} consensus algorithm. This protocol is not suitable for resource-constrained IoT devices like smart meters. Similarly, Li et al.~\cite{li2017consortium} suggest an ecosystem where prosumers can trade electricity using the blockchain-based medium of exchange, NRGcoin, which also employs \gls{PoW}. Although the consensus algorithm is executed on aggregators with enhanced processing power, it still contradicts the principle of energy efficiency.

Gai et al.~\cite{gai2019privacy} introduce a solution to ensure consumer-privacy regarding trading habits and patterns on the blockchain. However, this solution relies on a centralized exchange rather than a peer-to-peer bartering process, making it susceptible to a single point of failure risk. All trading activities pass through the Token Bank responsible for the transactions.

In contrast, \toolname addresses these challenges by utilizing a decentralized architecture, low-cost transactions, and a \gls{PoA} consensus algorithm, ensuring energy efficiency, scalability, and resilience against single point of failure.

\section{Conclusions}
\label{sec:conclusion}

Smart grids are experiencing rapid global adoption, but there is still considerable room for improvement in terms of performance, security, and efficacy. In this context, \toolname emerges as a groundbreaking solution that addresses the diverse needs and requirements of modern smart grids. By integrating the PKI/WoT architectures with blockchain technology, \toolname establishes a resilient infrastructure that is immune to CA-oriented threats, operating in a decentralized and distributed manner.

The scalability and resilience of \toolname were extensively validated through simulations on a virtualized testbed environment consisting of 500 nodes. The results demonstrate that \toolname outperforms the Web of Trust scheme, significantly reducing the time required for new nodes to establish trust relationships. Furthermore, even in the presence of compromised sections within the grid, \toolname ensures uninterrupted network operations, showcasing its robustness. Additionally, this paper presents a novel energy trading scheme based on smart contracts, leveraging the established trust between network nodes and adapting electricity prices based on factors such as distance to prosumers and the chosen production method. This secure energy trading mechanism plays a crucial role in unlocking the full potential of renewable energy sources.

With its comprehensive capabilities, \toolname has the potential to set the course for smart grids, unlocking their true potential and providing a solid foundation for the implementation of intelligent programs with diverse functionalities. As an ongoing research initiative, future work will apply the trust management scheme to a realistic testbed involving an actual smart grid deployment. Furthermore, \toolname will be enhanced with additional applications beyond energy trading, further cementing its position as a pioneering solution.

In summary, \toolname represents a significant advancement in revolutionizing smart grids, offering a comprehensive solution that addresses their current limitations. Its innovative design and proven performance make it a milestone achievement, propelling the industry towards a future where intelligent programs and cutting-edge functionalities find abundant opportunities for application.

\section*{Acknowledgements}
The authors would like to thank Dr. Apostolos Zarras for his valuable input in improving the readability and quality of this paper. His expertise and suggestions have greatly enhanced the overall clarity and flow of the manuscript. We are grateful for his valuable contributions and support throughout this research project.

\bibliographystyle{achemso}
\bibliography{references}
\end{document}